\documentclass[aps,pre,twocolumn,superscriptaddress,showpacs,showkeys,amsmath,amssymb]{revtex4}

\usepackage{epsfig,amsmath,amssymb,color}
\usepackage[T1]{fontenc}
\renewcommand{\theequation}{\arabic{section}.\arabic{equation}}

\newcommand{\myfigsize}{74mm}

\begin{document}

\title{Spin-1/2 $XX$ chain in a transverse field with regularly alternating $g$-factors:\\
       Static and dynamic properties}

\author{Taras Krokhmalskii}
\affiliation{Institute for Condensed Matter Physics,
          National Academy of Sciences of Ukraine,
          Svientsitskii Str. 1, 79011 L'viv, Ukraine}
\affiliation{Department for Theoretical Physics,
          Ivan Franko National University of L'viv,
          Drahomanov Str. 12, 79005 L'viv, Ukraine}

\author{Taras Verkholyak}
\affiliation{Institute for Condensed Matter Physics,
          National Academy of Sciences of Ukraine,
          Svientsitskii Str. 1, 79011 L'viv, Ukraine}

\author{Ostap Baran}
\affiliation{Institute for Condensed Matter Physics,
          National Academy of Sciences of Ukraine,
          Svientsitskii Str. 1, 79011 L'viv, Ukraine}

\author{Vadim Ohanyan}
\affiliation{Laboratory of Theoretical Physics,
          Yerevan State University,
          Alex Manoogian Str. 1, 0025 Yerevan, Armenia}
\affiliation{Joint Laboratory of Theoretical Physics -- ICTP Affiliated Centre in Armenia}
\affiliation{CANDLE Synchrotron Research Institute,
          Acharyan Str. 31, 0040 Yerevan, Armenia}

\author{Oleg Derzhko}
\affiliation{Institute for Condensed Matter Physics,
          National Academy of Sciences of Ukraine,
          Svientsitskii Str. 1, 79011 L'viv, Ukraine}
\affiliation{Department of Metal Physics,
          Ivan Franko National University of L'viv,
          Kyrylo \& Mephodiy Str. 8, 79005 L'viv, Ukraine}

\date{\today}

\begin{abstract}
We consider the spin-1/2 isotropic $XY$ chain in an external magnetic field directed along $z$ axis
with periodically varying $g$-factors.
To reveal the effects of regularly alternating $g$-factors,
we calculate various static and dynamic equilibrium quantities in the ground state and at finite temperatures.
We demonstrate that because of the regularly alternating $g$-factors
the saturation field may disappear
and
the field dependence of the susceptibility in the ground state has additional logarithmic singularity at zero field.
Moreover,
the zero-field susceptibility has a logarithmic singularity as $T\to 0$.
Furthermore,
the dynamic structure factors exhibit much more structure in the ``wave vector -- frequency'' plane
that can be traced out to modifications of the two-fermion excitation continua
which exclusively determine $S_{zz}(\kappa,\omega)$ and dominate the properties of $S_{xx}(\kappa,\omega)$.
We discuss what changes can be observed in dynamic experiments on the corresponding substances.
\end{abstract}



\maketitle

\section{Introduction}
\label{sec01}
\setcounter{equation}{0}

The magnetic moment of an electron is related to its angular momentum by the $g$-factor.
The magnetic moment of a free electron is associated with its spin angular moment only
and the magnitude of the electron $g$-factor (or more precisely the electron spin $g$-factor) is $\approx 2.002\,319$ \cite{Odom2007}.
In atoms, both orbital angular momentum and spin angular momentum of electron contribute to the magnetic moment of an atomic electron
and the spin $g$-factor has to be replaced by the Land\'{e} $g$-factor.
Furthermore,
in crystalline solids,
the Land\'{e} $g$-factor (or in what follows simply $g$-factor) may be, in principle, site dependent.

From the solid-state-physics side,
one can mention a number of spin-chain compounds
with regularly alternating $g$-factor values \cite{Niazi2002,Yin2013,Yin2015,Bhatt2014,CuMoDy,Kenzelmann2005,FeMnCu,Oshikawa1997,Coronado1986}.
Thus,
one-dimensional copper-iridium oxide Sr$_3$CuIrO$_6$
which contains both 3$d$ (Cu$^{2+}$) and 5$d$ (Ir$^{4+}$) magnetic ions
can be well described by an effective spin-1/2 ferromagnetic Heisenberg model
with an Ising-like exchange anisotropy ($\Delta\approx 2.5$) \cite{Niazi2002,Yin2013}.
Moreover,
the Cu sites carry the Cu spin $s=1/2$ with $g$-factor $\approx 2$
and
the Ir sites carry the Ir isospin $s=1/2$ with $g$-factor $\approx -3$ \cite{Yin2013,Yin2015}.
Another instance is a one-dimensional molecular magnet [\{Co$^{\rm{II}}(\Delta)$Co$^{\rm{II}}(\Lambda)$\}(ox)$_2$(phen)$_2$]$_n$ \cite{Bhatt2014}.
Magnetic properties of this compound can be explained using a one-dimensional Ising-chain model
with two different exchange couplings and two different $g$-factors, $2.5$ and $2.1$.
Next example of single-chain molecular magnet is a coordination polymer compound
$[\{(\text{CuL})_2\text{Dy}\}\{\text{Mo}(\text{CN})_8\}]\cdot\text{2CH}_3\text{CN}\cdot\text{H}_2\text{O}$,
in which L$^{2-}$ is N,N-propylenebis(3-methoxysalicylideneiminato).
The magnetic unit cell in this compound contains four magnetic ions with three different values of the $g$-factors.
The presence of highly anisotropic Dy$^{3+}$ ion makes possible an exact solution for the corresponding spin-chain model \cite{CuMoDy}.
One more example is the spin-1/2 chain antiferromagnet CuCl$_2\cdot$2((CD$_3$)$_2$SO) \cite{Kenzelmann2005}.
There are results of very recent studies of another heterotrimetallic coordination-polymer single-chain magnet
with large difference between the $g$-factors of the magnetic ions in the magnetic unit cell,
$[\text{Cu}^{II}\text{Mn}^{II}(\text{L}^{1})][\text{Fe}^{III}(\text{bpb})(\text{CN})_2]\cdot\text{ClO}_4\cdot\text{H}_2\text{O}$ \cite{FeMnCu}.
In this system,
a staggered $g$-tensor and/or Dzyaloshinskii-Moriya interactions
lead to a staggered field along $x$ direction upon application of a uniform field along $z$ direction.
As a result, a spin-1/2 antiferromagnetic Heisenberg chain with an alternating $g$-factor emerges
(see also Ref.~\cite{Oshikawa1997} discussing the quasi-one-dimensional spin-1/2 antiferromagnet Cu benzoate).
Finally,
one may also mention a two-sublattice one-dimensional system Ni$_2$(EDTA)(H$_2$O)$_4\cdot$2H$_2$O,
the magnetic behavior of which was discussed in terms of a spin-1 $g_1-g_2$ antiferromagnetic Heisenberg (or Ising) chain
with $g_1/g_2$ about $1.1$ \cite{Coronado1986}.

From the theoretical side,
since the $g$-factor enters many standard lattice models of crystalline solids,
it is quite natural to address a question
about the consequences of a regular non-uniformity of the $g$-factor for the observable magnetic properties.
There are several exact calculations for the spin-chain systems aimed on exploring the essential effects of nonuniform $g$-factors.
Spin-1/2 $XY$ chains provide an excellent playground for such analysis because they correspond to noninteracting fermions \cite{Lieb1961,Katsura1962}.
Prior work, which is closely related to our study, concerns
the two-sublattice \cite{Kontorovich1968,perk1975}
and
the inhomogeneous periodic
(i.e., with several sites in a cell which periodically repeats) \cite{Lima2006}
spin-1/2 $XX$ chain in a $z$-aligned field
with various interaction constant and $g$-factor values.
The reported results refer to the magnetization, susceptibility and equal-time two-spin $zz$ correlation functions \cite{Kontorovich1968,perk1975},
as well as to some dynamic quantities related to correlations of the average cell operators \cite{Lima2006}.
The continued-fraction method was also used to figure out the magneto-thermal properties of the general inhomogeneous isotropic $XX$ chain
including the case of  random Lorentzian transverse field \cite{Zaburannyi2000}.
The same program has been performed also for the quantum Ising chain \cite{quantIsing1}.
In the most recent papers,
the detailed analysis of the ground-state properties
for general boundary conditions for the quantum Ising chain with the period-2 modulated transverse field
have been done \cite{quantIsing2}.
Free-fermion models
in which the period-2 alternation of the nearest-neighbor interactions is accompanied by multiple spin exchange
were considered in Refs.~\cite{Zvyagin2006,Zvyagin2010,Zvyagin2016}.
$XX$ chains is the extreme limit of the Heisenberg chains with an $XY$-like exchange anisotropy.
The opposite limiting case is the Ising chains.
Recently,
a spin-1/2 Ising chain with period-2 regularly alternating $g$-factors has been studied
in context of unusual properties of Sr$_3$CuIrO$_6$ \cite{Yin2013,Yin2015}.
Moreover, this material, as was mentioned above, features not only alternating $g$-factors of magnetic ion along the chain,
but also the negative sign of the one of them.
Negative $g$-factors (for the pseudospin operators) are interesting by themselves
as they are the result of strong interplay between the ligand field and spin-orbit interaction \cite{ata08,chi12,chi13}.
Very recently it has been shown
that even in the simplest case of ferromagnetic Ising model with $g$-factors of different sign on bipartite lattice,
the frustration takes place and there are configurations containing ordered and disordered sublattices at the same time \cite{Yin2015,tor18}.
Rigorous results for finite quantum spin clusters and an Ising-Heisenberg chain with different $g$-factors
have been obtained recently in Ref.~\cite{Ohanyan2015}.

In the present paper we report results of the systematic study
of the spin-1/2 $XX$ chain in a transverse field with regularly alternating $g$-factors
including the case when $g$-factors have different sings.
We pay special attention to manifestation of regularly alternating $g$-factors
in the transverse magnetization, the static $zz$ susceptibility,
as well as
in the two dynamic structure factors $S_{zz}(\kappa,\omega)$ and $S_{xx}(\kappa,\omega)$.
$S_{yy}(\kappa,\omega)$ behaves identically to $S_{xx}(\kappa,\omega)$ due to the symmetry of the model.
Dynamic quantities are accessible experimentally
and therefore understanding of the effects generated by nonuniform $g$-factors
may be useful for interpreting experimental data.
The recent development of the exact and numerical calculations
of the spin dynamic structure factors for the integrable one-dimensional quantum spin systems are really impressive \cite{caux}.
However,
the examined in what follows spin-chain model,
although corresponds to noninteracting fermions,
may be of interest for the full Heisenberg exchange interaction case too:
Since the seminal papers by G.~M\"{u}ller et al. \cite{Mueller1981} we know
that many dynamic features of the spin-1/2 Heisenberg chain can be analyzed starting from the free-fermion limit.

It might be worth it to list here the main findings of the present paper.
\begin{itemize}
\item We have performed the detailed study of the dynamic properties.
We calculated the dynamic structure factors $S_{zz}(\kappa,\omega)$ and $S_{xx}(\kappa,\omega)$
and inspected how they change in the external magnetic field for different period-2 alternations of $g$-factors.
\item In the case when both $g$-factors are of the same sign,
the correspondence between the boundaries of the $zz$ and $xx$ structure factors is still present.
\item On the contrary, if $g_1 g_2 \leq 0$,
a large enough magnetic field leads to the highly intense modes in the $xx$ structure factor.
\item Analyzing the absorption intensity $I_{\alpha}(\omega,h)$, we found that in the Voigt configuration ($\alpha=z$),
the model with uniform $g$-factors does not have any response.
In the case when $g_2$ differs from $g_1$, we obtain the nonzero contribution to the absorption intensity.
For sufficiently large frequencies $\omega>2|J|$ (where $J$ denotes the
 exchange coupling) the van Hove singularity
 arises at the magnetic field $h=\sqrt{\omega^2-4J^2}/|g_1-g_2|$.
\item In the Faraday configuration ($\alpha=x$),
the situation is a bit different.
The absorption spectra can be observed in the uniform case.
It shows a broad maximum at some resonance field.
The alternation of $g$-factor leads to the doubling of this resonance line.
\item Although in our study we focus on the exactly solvable $XX$ chain,
we know that such analysis of dynamics is useful for understanding a more realistic case of the Heisenberg chains.
Many qualitative features
(e.g., doubling of the resonance line)
of the absorption profiles can be found also in case of Heisenberg of $XXZ$-model with alternating $g$-factors.
\end{itemize}

The rest of the paper is organized as follows.
We begin with introducing the model to be studied
and the free-fermion representation of the model which emerges after applying the Jordan-Wigner transformation,
Sec.~\ref{sec02}.
After that we discuss the magnetization and the susceptibility in the ground state (Sec.~\ref{sec03})
and some finite-temperature quantities (Sec.~\ref{sec04}).
In Sec.~\ref{sec05} we examine the dynamic structure factors of the model.
We report the results for $S_{zz}(\kappa,\omega)$ obtained mainly analytically and for $S_{xx}(\kappa,\omega)$ obtained mainly numerically.
We conclude the paper with a summary, Sec.~\ref{sec06}.

\section{The model and its free-fermion representation}
\label{sec02}
\setcounter{equation}{0}

In the present study,
we consider the spin-1/2 isotropic $XY$ chain in a transverse
(i.e., aligned along $z$ axis)
magnetic field.
The peculiarity of the model is the regularly alternating $g$-factor
which acquires periodically two values, $g_1$ and $g_2$.
The Hamiltonian of the model reads
\begin{eqnarray}
\label{201}
H
&=&
\sum_{l=1}^{\frac{N}{2}}
\left[
J\left(s_{2l-1}^x s_{2l}^x + s_{2l-1}^y s_{2l}^y + s_{2l}^x s_{2l+1}^x + s_{2l}^y s_{2l+1}^y\right)
\right.
\nonumber\\
&&\left.
- g_1 \mu_{\rm{B}} {\sf{H}} s_{2l-1}^z - g_2 \mu_{\rm{B}} {\sf{H}} s_{2l}^z
\right].
\end{eqnarray}
Here $J$ is the exchange interaction
(we may put $\vert J\vert=1$ without loss of generality),
$\mu_{{\rm{B}}}$ is the Bohr magneton,
${\sf{H}}$ is the value of the magnetic field
measured, e.g., in Teslas
(then with $\mu_{{\rm{B}}}\approx 0.67171$K/T the field $h=\mu_{{\rm{B}}}{\sf{H}}$ is measured in Kelvins),
and
$g_1\mu_{{\rm{B}}}{\sf{H}}=g_1h$,
$g_2\mu_{{\rm{B}}}{\sf{H}}=g_2h$.
Furthermore,
$N$ is the number of lattice sites which is assumed to be even,
and periodic boundary conditions are imposed for convenience.
After introducing
\begin{eqnarray}
\label{202}
g_{\pm}=\frac{g_1\pm g_2}{2},
\end{eqnarray}
we can rewrite Eq.~(\ref{201}) in a more compact form
\begin{eqnarray}
\label{203}
&&H=\sum_{l=1}^N\left[J\left(s_l^x s_{l+1}^x + s_l^y s_{l+1}^y\right) - h_l s_{l}^z\right],
\nonumber\\
&&h_l= [g_+ - (-1)^lg_-]h.
\end{eqnarray}
This is the Hamiltonian of the spin-1/2 isotropic $XY$ chain in a regularly alternating (with period 2) transverse magnetic field.

The defined model is exactly solvable by making use of the famous Jordan-Wigner fermionization \cite{Lieb1961,Katsura1962}
(see also Refs.~\cite{Derzhko2001,Derzhko2008}).
In terms of the Jordan-Wigner fermions
the spin Hamiltonian (\ref{203}) becomes
\begin{eqnarray}
\label{204}
H=\sum_{l=1}^N\left[\frac{J}{2}\left(c_l^\dagger c_{l+1} + c_{l+1}^\dagger c_l \right)
-h_l\left(c_l^\dagger c_l -\frac{1}{2}\right)\right].
\end{eqnarray}
Again periodic boundary conditions are implied in Eq.~(\ref{204}) \cite{note1}.
After the Fourier transformation
\begin{eqnarray}
\label{205}
&&c_l=\frac{1}{\sqrt{N}}\sum_\kappa e^{-i\kappa l} c_\kappa,
\nonumber\\
&&\kappa=\frac{2\pi j}{N}, \quad
j=-\frac{N}{2},-\frac{N}{2}+1, \ldots ,\frac{N}{2}-1,
\end{eqnarray}
Eq.~(\ref{204}) can be cast into
\begin{eqnarray}
\label{206}
H&=&\sum_{-\pi\leq\kappa<\pi}\left[(J\cos\kappa - g_+h) c_{\kappa}^\dagger c_{\kappa} +g_-hc_{\kappa}^\dagger c_{\kappa\pm\pi}\right]
\nonumber \\
&+& \frac{g_+h}{2}N.
\end{eqnarray}
Next, we perform the Bogolyubov transformation,
\begin{eqnarray}
\label{207}
&&c_{\kappa}= u_{\kappa}\alpha_{\kappa} - v_{\kappa}\alpha_{\kappa+\pi},
\\ \nonumber
&&c_{\kappa+\pi}= v_{\kappa}\alpha_{\kappa} + u_{\kappa}\alpha_{\kappa+\pi} \quad
(-{\pi}/{2} \leq \kappa < {\pi}/{2}),
\\
&&u_{\kappa}=\frac{1}{\sqrt 2}\sqrt{1+\frac{|J\cos\kappa|}{\sqrt{J^2\cos^2\kappa+g_{-}^2h^2}}},\nonumber\\
&&v_{\kappa}=\frac{{\rm{sgn}}(g^{}_{-}h J\cos\kappa)}{\sqrt 2}\sqrt{1-\frac{|J\cos\kappa|}{\sqrt{J^2\cos^2\kappa+g_{-}^2h^2}}},\nonumber
 \end{eqnarray}
leading to
\begin{eqnarray}
\label{208}
&&H=
\sum_{-\pi\leq\kappa<\pi}\Lambda_\kappa \left(\alpha_\kappa^\dagger \alpha_\kappa -\frac{1}{2}\right),
\\
&&\Lambda_\kappa
=
-g_+h +{\rm{sgn}}(J\cos\kappa)\sqrt{J^2\cos^2\kappa+g_-^2h^2}.\nonumber
\end{eqnarray}
Hence,
we have arrived at the free-fermion representation \eqref{208} of the initial spin model (\ref{201}).
Within this representation
many calculations for the thermodynamically large system can be performed rigorously analytically or with very high accuracy numerically.
From Eq.~(2.8) it is immediately evident
that nonzero magnetic field develops a gap in the excitation spectrum splitting it into two branches.
In the limiting case of large $g$-factors (or field $h$)
the system becomes close to the two-level model with only two possible eigenenergies on each site $-g_1h$ and $-g_2h$.
The position of the Fermi level is important for the understanding of the ground state and thermodynamics of the model given in the next section.

Although the isotropic $XY$ interactions may occur in some spin-1/2 chain compounds
(see, e.g., Ref.~\cite{Kenzelmann2002}),
they can be viewed as a limiting case of more common $XXZ$ interactions.
Consider the spin-1/2 $XXZ$ chain in a $z$-directed magnetic field.
The Hamiltonian of such model contains
in addition to the one given in Eqs.~(\ref{201}) or (\ref{202})
the interaction of the $z$ components of neighboring spins with the strength $J\Delta$,
where $\Delta$ is the anisotropy parameter.
As a result, in terms of the Jordan-Wigner fermions the spin Hamiltonian becomes
\begin{eqnarray}
H&=&\sum_{l=1}^N
\left[
\frac{J}{2}\left(c_l^\dagger c_{l+1}+c_{l+1}^\dagger c_l\right)
+J\Delta c_l^\dagger c_l c_{l+1}^\dagger c_{l+1}
\right.\nonumber\\
&-&\left.
\left(h_l +J\Delta\right)c_l^\dagger c_l+\frac{h_l}{2}+\frac{J\Delta}{4}
\right].
\end{eqnarray}
One way to proceed is to apply a mean-field like approximation for the four-fermion term \cite{Bulaevskii1963,Zvyagin2020}:
\begin{eqnarray}
c_l^\dagger c_l c_{l+1}^\dagger c_{l+1}
&\rightarrow&
\left(\frac{1}{2}{+}m\right)\left(c_l^\dagger c_l {+} c_{l+1}^\dagger c_{l+1}\right) {-} \left(\frac{1}{2}{+}m\right)^2
\nonumber\\
&&-t \left(c_l^\dagger c_{l+1} + c_{l+1}^\dagger c_{l}\right)+t^2
\nonumber\\
&&-s c_l^\dagger c_{l+1}^\dagger - s^* c_l c_{l+1} +\vert s\vert^2,
\end{eqnarray}
where the parameters
$m\equiv \langle c_l^\dagger c_l\rangle -1/2$,
$t\equiv \langle c_l^\dagger c_{l+1}\rangle$,
and
$s\equiv \langle c_l c_{l+1}\rangle$
have to be determined self-consistently.
It should be noted that the Jordan-Wigner fermionization approach was successfully used
for examining the static and dynamic properties away from the free-fermion point \cite{Caux2003,Dmitriev2002,Hagemans2005,Nunner2004,Bruognolo2016}.

\section{Zero-temperature properties}
\label{sec03}
\setcounter{equation}{0}

Let us first present the ground-state ($T=0$) properties of the system.
Although some particular results have been already obtained in Refs.~\cite{Kontorovich1968,perk1975,Lima2006},
we provide here the ground-state analysis for consistency.
Particularly, we focus on calculating
the ground-state energy $e_0=\langle H \rangle/N$,
the transverse magnetization $m=-\partial e_0/\partial h$,
the sublattice average $z$-component of spin,
$\langle s_1^z \rangle=-2 \, \partial e_0/\partial (g_1h)$, $\langle s_2^z \rangle=-2 \, \partial e_0/\partial (g_2h)$,
and the static $zz$ susceptibility $\chi_{zz}=\partial m/\partial h$.
For the model at hand, one has to differ the magnetization and the average of the $z$-component of the spin operator,
i.e., the magnetic moment and the angular moment at site.
It is obvious that
\begin{eqnarray}
\label{301}
m=\frac 12\left(g_1\langle s_1^z \rangle+g_2\langle s_2^z \rangle\right).
\end{eqnarray}
In what follows we distinguish two cases: $g_1g_2>0$ and $g_1g_2<0$.

\underline{The case $g_1g_2>0$.}
There are two values of the Fermi momenta $\kappa_F$
defined as the solutions of the equation $\Lambda_{\kappa}=0$:
\begin{eqnarray}
\label{302}
&& \kappa_F=\pm\kappa_0 , \;\; {\rm if} \;\;
0<Jg_+h<|Jg_+|h_s,
\\
&& \kappa_F=\pm(\pi - \kappa_0), \;\; {\rm if} \;\;
-|Jg_+|h_s<Jg_+h<0,
\nonumber\\
&& \kappa_0=\arccos\left|{h}/{h_s}\right| \quad (0<\kappa_0<{\pi}/{2}),
\nonumber
\end{eqnarray}
where the saturation field $h_s$ is given by $h_s=\vert J\vert/\sqrt{g_1g_2}>0$.
It is worth to note that the saturation field exists
if the fully polarized state $\vert \uparrow\ldots\uparrow\rangle$,
which is obviously the eigenstate of the Hamiltonian (2.1),
becomes the ground state as the field $h$ exceeds a certain finite value.
This is the case for $g_1g_2>0$ but not for $g_1g_2<0$.
Here we may consider two separate ranges of the magnetic field $h$. The first one, when $|h|> h_s$, corresponds to the saturated phase with all spins aligned in the field direction. There is no solution for $\kappa_F$ and, thus, the ground state energy as well as the averages of spins have simple expressions:
\begin{eqnarray}
\label{303}
&& e_0=-\frac 12|g_+h|,\quad m={\rm{sgn}}(h)\frac{g_+}{2},
\\
&&\langle s_1^z \rangle=\langle s_2^z \rangle=\frac{{\rm{sgn}}(h)}{2},\quad \chi_{zz}=0.
\nonumber
\end{eqnarray}
More interesting is the second range, $-h_s<h<h_s$, when
\begin{eqnarray}
\label{304}
&&e_0={-}|g_+^{}h|\left(\frac{1}{2}{-}\frac{\kappa_0}{\pi}\right)
{-}\frac{1}{\pi}\sqrt{J^2+g_-^2h^2}{\rm{E}}(\kappa_0,\varkappa),
\\
&&m=g_+^{}{\rm{sgn}}(h)\left(\frac{1}{2}-\frac{\kappa_0}{\pi}\right)
+\frac{g_-^2 h}{\pi\sqrt{J^2+g_-^2h^2}}{\rm{F}}(\kappa_0,\varkappa),
\nonumber\\
&&\langle s_1^z\rangle={\rm{sgn}}(h)\left(\frac{1}{2}-\frac{\kappa_0}{\pi}\right)+\frac{g_-^{}h}{\pi\sqrt{J^2+g_-^2h^2}}{\rm{F}}(\kappa_0,\varkappa),
\nonumber\\
&&\langle s_2^z\rangle={\rm{sgn}}(h)\left(\frac{1}{2}-\frac{\kappa_0}{\pi}\right)-\frac{g_-^{}h}{\pi\sqrt{J^2+g_-^2h^2}}{\rm{F}}(\kappa_0,\varkappa),
\nonumber\\
&&\chi_{zz}=\frac{g_+^{} \varkappa^2}{\pi\sqrt{h_s^2{-}h^2}}{+}\frac{g_-^2}{\pi\sqrt{J^2{+}g_-^2h^2}}
\left({\rm{F}}(\kappa_0,\varkappa){-}{\rm{E}}(\kappa_0,\varkappa)\right).
\nonumber
\end{eqnarray}
Here $\varkappa=\vert J\vert/\sqrt{J^2+g_-^2h^2}$
and we have also introduced the elliptic integrals of the first and second kind given by the following standard expressions \cite{Jahnke}:
\begin{eqnarray}
\label{305}
&&{\rm{F}} (\kappa_0,\varkappa )
=\int_0^{\kappa_0}\frac{{\rm{d}}\theta}{\sqrt{1-\varkappa^2\sin^2\theta}},
\\
&&{\rm{K}} (\varkappa )={\rm{F}} \left(\frac{\pi}{2},\varkappa \right),
\nonumber\\
&&{\rm{E}} (\kappa_0,\varkappa )
=\int_0^{\kappa_0}{\rm{d}}\theta \sqrt{1-\varkappa^2\sin^2\theta},
\nonumber\\
&&{\rm{E}} (\varkappa )={\rm{E}} \left(\frac{\pi}{2},\varkappa \right).
\nonumber
\end{eqnarray}

As can be seen from the reported formulas,
the susceptibility diverges at $h=\pm h_s$ showing the square-root singularity
\begin{eqnarray}
\label{306}
\chi_{zz}\approx
\frac{g_{+}^2-g_{-}^2}{\pi g_+}\frac{1}{\sqrt{h_s^2-h^2}},
\,\,\,
h\to\vert h_s\vert.
\end{eqnarray}
If $g_1\ne g_2$ an additional weak divergence of $\chi_{zz}$ occurs at $h=0$:
\begin{eqnarray}
\label{307}
\chi_{zz}\approx
\frac{g_+}{\pi h_s}
+\frac{g_-^2}{\pi}
\left(\ln\frac{2h_s}{h} - 1\right),
\,\,\,
\vert h\vert\to 0.
\end{eqnarray}
It was noticed for the first time apparently in Ref.~\cite{Kontorovich1968}.

\underline{The case $g_1g_2<0$.}
In this case the equation for the Fermi momenta $\Lambda_{\kappa}=0$ does not have real solutions,
which means that the Fermi level lays in the forbidden band between two branches of the spectrum.
Since the odd and even spins are directed oppositely in a field,
there is also no saturation field,
i.e., the magnetization never attains its saturation value corresponding to $\langle s_1^z\rangle=-\langle s_2^z\rangle=\pm 1/2$.
The ground-state energy is given by the following formula:
\begin{eqnarray}
\label{308}
e_0=-\frac{1}{\pi}\sqrt{J^2+g_-^2h^2}{\rm{E}}(\varkappa).
\end{eqnarray}
After straightforward differentiation we get
\begin{eqnarray}
\label{309}
&&m=\frac{g_-^2 h}{\pi\sqrt{J^2+g_-^2h^2}} {\rm{K}}(\varkappa),
\\
&&\langle s_1^z\rangle=-\langle s_2^z\rangle=\frac{g_-h}{\pi\sqrt{J^2+g_-^2h^2}}{\rm{K}}(\varkappa),
\nonumber\\
&&\chi_{zz}=\frac{g_-^2}{\pi\sqrt{J^2+g_-^2h^2}}\left({\rm{K}}(\varkappa)-{\rm{E}}(\varkappa)\right)
\nonumber
\end{eqnarray}
for the magnetization, the sublattice average $z$-component of spin, and the susceptibility, respectively.
These formulas can be simplified in the strong-field and weak-field limits.
We obtain
\begin{eqnarray}
\label{310}
&&m\approx\frac{g_-^2h}{2\sqrt{J^2+g_-^2h^2}},
\\
&&\chi_{zz}\approx\frac{g_-^2 J^2}{4\left(J^2+g_-^2h^2\right)^{\frac{3}{2}}},
\nonumber
\end{eqnarray}
as $\vert h\vert\to\infty$
and
\begin{eqnarray}
\label{311}
&&m\approx
\frac{g_-^2h}{\pi\sqrt{J^2+g_-^2h^2}}
\ln\frac{4\sqrt{J^2+g_-^2h^2}}{\vert g_-h\vert},
\\
&&\chi_{zz}\approx
\frac{g_-^2}{\pi\sqrt{J^2+g_-^2h^2}}
\left(\ln\frac{4\sqrt{J^2+g_-^2h^2}}{\vert g_-h\vert}-1\right),
\nonumber
\end{eqnarray}
as $\vert h\vert\to 0$.
While Eq.~(\ref{310}) demonstrates explicitly that the saturation is never achieved for any finite $h$,
Eq.~(\ref{311}) demonstrates a non-analyticity of the ground-state energy
which manifests itself as a logarithmic peculiarity of the magnetization and the susceptibility in vanishing field.

In Fig.~1 we show the ground-state magnetization and susceptibility.
In all numerical investigations, without loss of generality, we assume first that $g_2=g_1=1$ and then $g_2$ starts to decrease.
These plots illustrate the reported above analytical results including the asymptotic behavior of the susceptibility.
It is worthwhile to stress
that the logarithmic singularity of the susceptibility $\chi_{zz}$ can be detected not only in the case $g_1g_2<0$,
when it is quite natural to expect it,
but also in the opposite case $g_1g_2>0$, see Eq.~(3.7).
It is the consequence of another peculiar property shown in Fig.~2
where the total magnetization and spin moment is confronted with the average spin moments of each sublattices.
We can see that even for positive $g_2$ (see Fig.~2 for $g_2=0.1$)
the average spin moment at small fields started to evolve in the opposite to the field direction
feeling the competition between the applied magnetic field and quantum interaction with stronger magnetized neighboring spins.

Let us denote by $h_0$ ($h_0>0$) the value of the field
at which $\langle s_2^z\rangle=0$
if $|g_2|<|g_1|$ (or $\langle s_1^z\rangle=0$ if $|g_1|<|g_2|$);
$h_0$ exists in the case $g_1g_2>0$ only.
After using approximate formulas for the elliptic integrals one can show that $h_0\approx 2h_s \, e^{-2\alpha}$,
where $\alpha=\sqrt{g_1g_2}/\vert g_1-g_2\vert$.
If $g_2$ (or $g_1$) approaches zero
we can again use approximate formulas for the elliptic integrals to conclude that $h_0\approx h_s/\sqrt{2}$.
Both limiting cases can be combined into the following approximate expression
\begin{eqnarray}
\label{312}
h_0\approx\frac{2e^{-2\alpha}}{1+(2\sqrt{2}-1)e^{-3\alpha}}h_s,
\end{eqnarray}
which yields the correct value of $h_0$ for the whole region $g_1g_2>0$ with the accuracy of less than 1.5\%.

\begin{figure}
\begin{center}
\includegraphics[clip=on,width=\myfigsize]{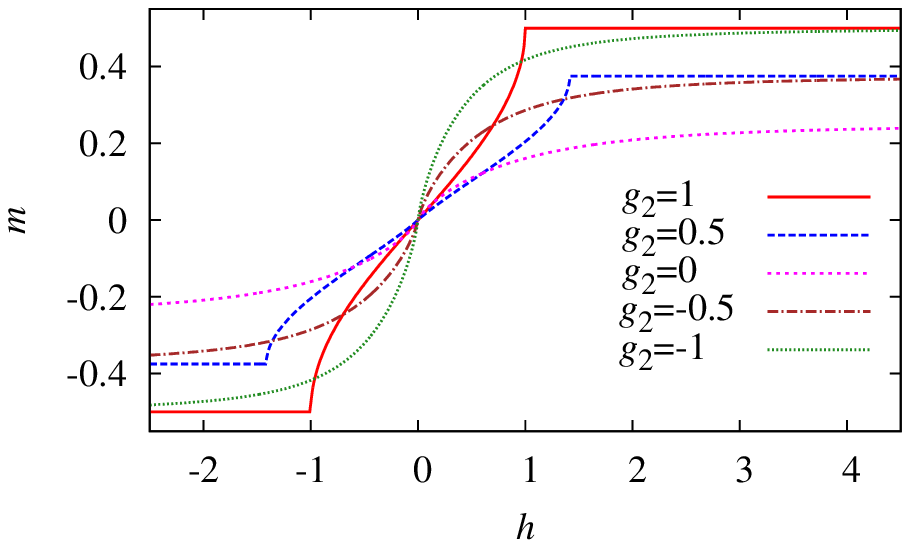}
\includegraphics[clip=on,width=\myfigsize]{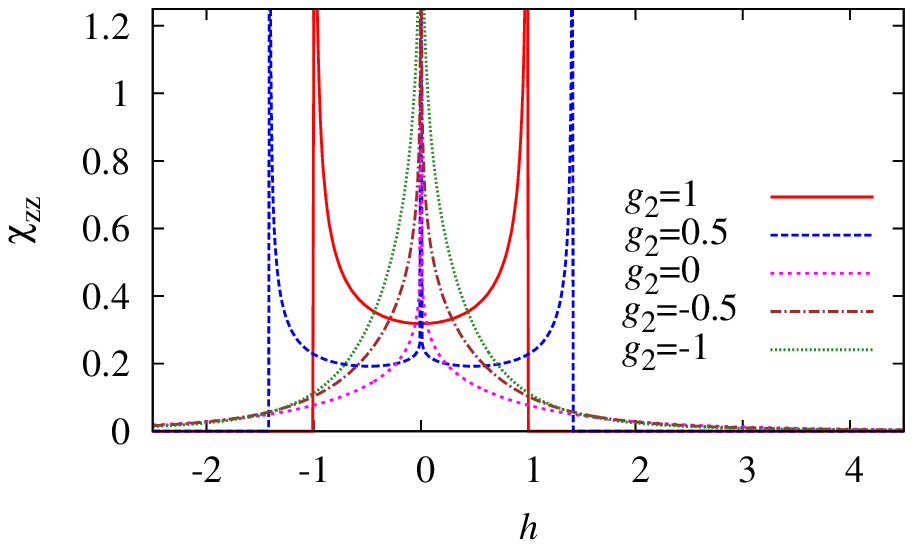}
\caption
{(Color online)
Ground-state magnetization (upper panel) and susceptibility (lower panel) vs field $h$.
$|J|=1$, $g_1=1$,
$g_2=1$ (solid),
$g_2=0.5$ (long-dashed),
$g_2=0$ (short-dashed),
$g_2=-0.5$ (dashed-dotted),
$g_2=-1$ (dotted).}
\label{x01}
\end{center}
\end{figure}

\begin{figure}
\begin{center}
\includegraphics[width=\myfigsize]{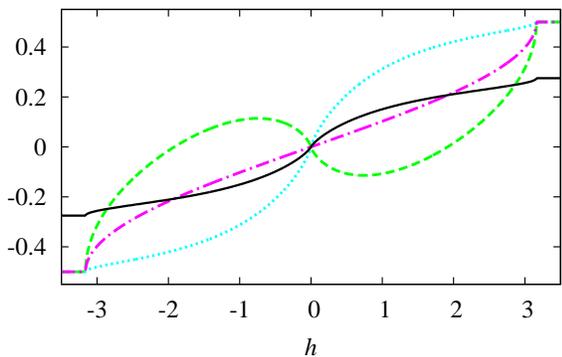}
\caption
{(Color online)
Ground-state values of
$\langle s_1^z\rangle$ (dotted),
$\langle s_2^z\rangle$ (dashed),
$(\langle s_1^z\rangle+\langle s_2^z\rangle)/2$ (dot-dashed),
and
$m$ (solid)
vs field $h$.
$|J|=1$, $g_1=1$, $g_2=0.1$.}
\label{x02}
\end{center}
\end{figure}

\section{Finite-temperature properties}
\label{sec04}
\setcounter{equation}{0}

Finite-temperature quantities can be easily calculated from the free energy per site
\begin{eqnarray}
\label{401}
f(T,h)
=
-\frac{T}{2\pi}\int_{-\pi}^{\pi}{\mbox{d}}\kappa\ln\left(2\cosh\frac{\Lambda_\kappa}{2T}\right)
\end{eqnarray}
with $\Lambda_\kappa$ given in Eq.~(\ref{208}).
For example,
for the specific heat one finds
\begin{eqnarray}
\label{402}
c(T,h)
=\frac{1}{2\pi}\int_{-\pi}^{\pi}{\mbox{d}}\kappa \left(\frac{\Lambda_\kappa}{2T}\right)^2\cosh^{-2}\frac{\Lambda_\kappa}{2T}.
\end{eqnarray}
Furthermore,
for the finite-temperature magnetization and susceptibility one finds
\begin{eqnarray}
\label{403}
m(T,h)=
\frac{1}{4\pi}\int_{-\pi}^{\pi}{\mbox{d}}\kappa\frac{\partial \Lambda_\kappa}{\partial h}
\tanh\frac{\Lambda_\kappa}{2T}
\end{eqnarray}
and
\begin{eqnarray}
\label{404}
&&\chi_{zz}(T,h)=
\\
&&\frac{1}{4\pi}\int_{-\pi}^{\pi}{\mbox{d}}\kappa
\left[\frac{\partial^2\Lambda_\kappa}{\partial h^2}\tanh\frac{\Lambda_\kappa}{2T}
+\frac{1}{2 T}\left(\frac{\partial\Lambda_\kappa}{\partial h}\right)^2\cosh^{-2}\frac{\Lambda_\kappa}{2T}\right],
\nonumber
\end{eqnarray}
respectively.
Here, the derivatives $\partial\Lambda_\kappa/\partial h$ and $\partial^2\Lambda_\kappa/\partial h^2$ are given by the following formulas:
\begin{eqnarray}
\label{405}
&&\frac{\partial\Lambda_\kappa}{\partial h}=-g_+ +\frac{{\rm{sgn}}(J\cos\kappa) g_-^2 h}{\sqrt{J^2\cos^2\kappa+g_-^2h^2}},
\\
&&\frac{\partial^2\Lambda_\kappa}{\partial h^2}= \frac{{\rm{sgn}}(J\cos\kappa)g_-^2J^2\cos^2\kappa}{\left(J^2\cos^2\kappa+g_-^2h^2\right)^{3/2}}.
\nonumber
\end{eqnarray}

\begin{figure}
\begin{center}
\includegraphics[width=\myfigsize]{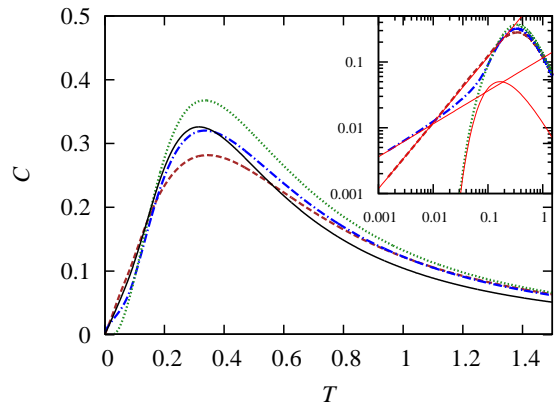}
\caption
{(Color online)
Temperature dependence of the specific heat for $|J|=1$ at
$h=0$ (solid black);
$h=0.5, g_1=1, g_2=0.5$ (dashed brown);
$h=0.5, g_1=1, g_2=0$ (dashed-dotted blue);
and
$h=0.5, g_1=1, g_2=-0.5$ (dotted green).
The inset shows the same plots in $\log-\log$ scale.
The linear, square-root and exponential behavior of the specific heat are clearly visible here.
Thin red lines represent the asymptotic forms from Eqs.~(\ref{406}), (\ref{407}), and (\ref{408}).}
\label{x03}
\end{center}
\end{figure}

In Fig.~\ref{x03} we demonstrate the temperature behavior of the specific heat \eqref{402} for several regimes:
1) gapless zero-field and finite-field regimes ($0<|h| <h_s$)
(solid black and dashed brown),
2) two cases when $|h|=h_s$ or $g_2=0$
(dashed-dotted blue),
and
3) two gapped regimes when $|h|>h_s$, $g_1g_2>0$ or when $g_1g_2<0$ at $h\ne0$
(dotted green).

The gapless regime features the universal linear-temperature dependence of the specific heat:
\begin{eqnarray}
\label{406}
c(T)\simeq \frac{\pi {\sf c}}{3 v_F}T,
\;\;\;
T\to 0.
\end{eqnarray}
Here, in our case
the central charge ${\sf c}=1$
and
the Fermi velocity for the case of zero field coincides with the those for the $XX$-chain, $v_F=|J|$,
whereas
for the case of the gapless finite-field regime ($0<|h|<h_s$, $g_1g_2>0$) it is $v_F=J^2\sqrt{1-h^2/h_s^2}/(h_s |g_+|)$.
When the magnetic field reaches the saturation value $|h|=h_s$ ($g_1g_2>0$)
the Fermi level touches the bottom points of the upper part of the spectrum (van Hove singularity).
The low-temperature behavior of the specific heat in this case is given by the square-root temperature dependence,
\begin{eqnarray}
\label{407}
c(T)\simeq\frac {3\left(\sqrt 2-1\right)\zeta\left(\frac 32\right) \sqrt{|g_+h|}}{8 \sqrt{\pi} |J|}\sqrt{T},
\end{eqnarray}
where $\zeta\left(x\right)$ is the standard zeta-function.
The same expression is valid for the case $g_2=0$ for arbitrary nonzero values of the magnetic field.
Finally, two gapped regimes are possible:
i) $|h|>h_s$, $g_1g_2>0$
and
ii) $g_1g_2<0$ at any $h\neq 0$.
The specific heat has universal exponential low-temperature behavior, given by
\begin{eqnarray}
\label{408}
c(T)\simeq\frac{\Delta^2 }{\sqrt{2\pi r}}\frac{e^{-\frac{\Delta}{T}}}{T^{\frac{3}{2}}},
\end{eqnarray}
where
for the $|h|>h_s$ regime $r=|J|\varkappa/2$, $\Delta=|g_+h|-\sqrt{J^2+g_-^2h^2}$,
whereas
for the $g_1 g_2<0$ regime $r=J^2/(2|g_-h|)$, $\Delta=|g_2 h|$ ($\Delta=|g_1 h|$) if $|g_2|<|g_1|$ ($|g_2|>|g_1|$).

\begin{figure}
\begin{center}
\includegraphics[width=\myfigsize]{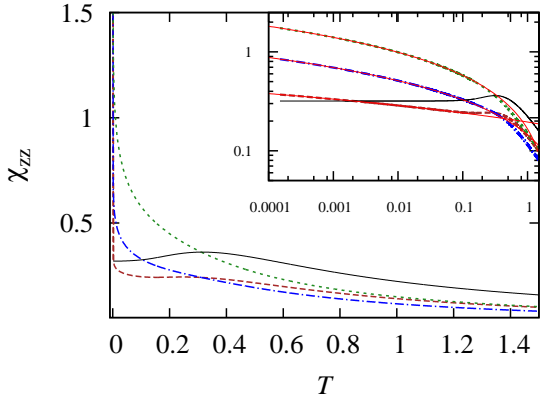}
\caption
{(Color online)
Low-temperature behavior of the zero-field susceptibility for $|J|=1$, $g_1=1$ and
$g_2=1$ (solid black),
$g_2=0.5$ (dashed brown),
$g_2=0$ (dashed-dotted blue),
and
$g_2=-0.5$ (dotted green).
The inset shows the same plots in $\log-\log$ scale.
Thin red lines represent the asymptotic form from Eq.~(\ref{chi_h=0}).}
\label{x04}
\end{center}
\end{figure}

Let us also consider the low-temperature behavior of the magnetic susceptibility at zero field.
We have the universal formula with logarithmic singularity given by
\begin{eqnarray}
\label{chi_h=0}
\chi_{zz}(T)\simeq\frac{1}{\pi|J|}\left[g_+^2-g_-^2\left(\ln\frac{\pi T}{4|J|}-\mathcal{C}\right)\right],
\end{eqnarray}
where $\mathcal{C}\simeq0.577\,215\,6$ is the Euler-Mascheroni constant.
As it is seen from this expression,
the logarithmic divergence at $T\rightarrow0$ is the consequence of the non-uniformity of the $g$-factors
and it disappears when $g_-=0$.
This is illustrated in Fig.~\ref{x04}.

\section{Dynamic properties}
\label{sec05}
\setcounter{equation}{0}

In this section,
we study dynamic quantities of the model.
Dynamic properties of quantum spin-chain compounds are observable
in the neutron scattering \cite{zaliznyak2013} and electron spin resonance (ESR) \cite{Ajiro2003} experiments.

We start with the dynamic structure factor related to the inelastic neutron scattering cross section \cite{zaliznyak2013,Jensen}:
\begin{eqnarray}
\label{501}
&&S_{\alpha\alpha}(\kappa,\omega)
=\\
&&\frac{1}{N}
\sum_{j=1}^N\sum_{n=1}^{N}\exp\left({\rm{i}}\kappa n\right)
\int\limits_{-\infty}^{\infty}{\rm{d}}t\exp\left({\rm{i}}\omega t\right)
g_jg_{j+n}\langle s_j^\alpha (t) s_{j+n}^\alpha \rangle_c,
\nonumber
\end{eqnarray}
where
$\langle s_j^\alpha (t) s_{j+n}^\alpha \rangle_c
=\langle s_j^\alpha (t) s_{j+n}^\alpha \rangle - \langle s_j^\alpha \rangle\langle s_{j+n}^\alpha \rangle$
and
$s_j^\alpha (t) = \exp(iHt)s_j^\alpha \exp(-iHt)$.
The inclusion of the $g$-factors in Eq.~(\ref{501}) here implies that we have the dynamic structure factors of the magnetic moments.
In general, $g$-factors may also depend on the probing field direction $\alpha$.
But if we imply that the ratio between $g_1$ and $g_2$ is preserved for any direction $\alpha$, Eq.~(\ref{501}) will acquire a scaling factor.
In the case of site-independent $g$-factors Eq.~(\ref{501}) coincides with the definition of Refs.~\cite{Derzhko1997,Derzhko2000,Derzhko2002}.
For the chain with site-dependent $g$-factors with period two the dynamic structure factor has the following general structure:
\begin{eqnarray}
\label{502}
S_{\alpha\alpha}(\kappa,\omega)
&=&g_{+}^2S_{\alpha\alpha}^{0}(\kappa,\omega) + g_{-}^2S_{\alpha\alpha}^{0}(\kappa+\pi,\omega)
\\
&-&g_-g_+\left(\overline{S}_{\alpha\alpha}^{0}(\kappa,\omega)+\overline{S}_{\alpha\alpha}^{0}(\kappa+\pi,\omega)\right),
\nonumber
\end{eqnarray}
where
the uniform spin structure factor $S_{\alpha\alpha}^{0}(\kappa,\omega)$
and
the staggered spin structure factor $\overline{S}_{\alpha\alpha}^{0}(\kappa,\omega)$
are defined in the standard way:
\begin{eqnarray}
\label{503}
&&S_{\alpha\alpha}^{0}(\kappa,\omega)
=\\
&&\frac{1}{N}
\sum_{j=1}^N\sum_{n=1}^{N}\exp\left({\rm{i}}\kappa n\right)
\int\limits_{-\infty}^{\infty}{\rm{d}}t\exp\left({\rm{i}}\omega t\right)
\langle s_j^\alpha (t) s_{j+n}^\alpha \rangle_c,
\nonumber\\
&&\overline{S}_{\alpha\alpha}^{0}(\kappa,\omega)
=\nonumber\\
&&\frac{1}{N}
\sum_{j=1}^N\sum_{n=1}^{N}\exp\left({\rm{i}}\kappa n\right)
\int\limits_{-\infty}^{\infty}{\rm{d}}t\exp\left({\rm{i}}\omega t\right)
(-1)^j \langle s_j^\alpha (t) s_{j+n}^\alpha \rangle_c.
\nonumber
\end{eqnarray}

Furthermore,
we consider $S_{zz}(\kappa,\omega)$ and $S_{xx}(\kappa,\omega)$ structure factors
separately.
In the former case one faces a problem of two-fermion excitations only and all calculations can be performed analytically.
The latter case corresponds to many-fermion excitations problem and requires, in general, the calculation of Pfaffians.
We perform these calculations numerically \cite{Derzhko1997,Derzhko2000,Derzhko2002,Derzhko2008} carefully controlling the accuracy of computations.
As in previous studies on the dynamics of spin-1/2 $XY$ chains,
both structure factors exhibit some similarities.
In what follows, we discuss the changes in these quantities caused by regular alternation of $g$-factors.

The dynamic structure factors allow us to calculate the energy absorption intensities $I_{\alpha}(\omega,h)$, $\alpha=z,x$ observed in the ESR experiments.
Following the procedure given in Appendix A of Ref.~\cite{Brockmann2012}, we can get for the linearly polarized electromagnetic wave:
\begin{eqnarray}
\label{504}
I_\alpha(\omega,h)&\propto&\omega\chi''_{\alpha\alpha}(0,\omega),
\\
\chi''_{\alpha\alpha}(0,\omega)&=&\frac{1-\exp(-\beta\omega)}{2}S_{\alpha\alpha}(0,\omega),
\nonumber
\end{eqnarray}
where $\chi''_{\alpha\alpha}(0,\omega)$ is the imaginary part of the $\alpha\alpha$ dynamic susceptibility
and $S_{\alpha\alpha}(0,\omega)$ is the corresponding dynamic structure factor at $\kappa=0$ defined in Eq.~(\ref{501}).
In the ESR experiment two configurations are distinguished \cite{Ajiro2003}:
i) the Voigt configuration, when the magnetic polarization of the electromagnetic wave is collinear with the constant field,
and
ii) the Faraday configuration, when the magnetic polarization of the electromagnetic wave is perpendicular to the constant field.
In our model, the $z$ [$x$] polarized electromagnetic wave corresponds to the Voigt [Faraday] configuration,
i.e., the absorption intensity is $I_z(\omega,h)$ [$I_x(\omega,h)$].
Again, as discussed in what follows, the regularly alternating $g$-factors change dramatically the ESR absorption intensity.

\subsection{$zz$ dynamics}

One can work out the closed-form expression for the dynamic structure factor $S_{zz}(\kappa,\omega)$.
It is given by the following expression:
\begin{eqnarray}
\label{505}
&&S_{zz}(\kappa,\omega){=}
\int\limits_{-\pi}^\pi{\rm{d}}\kappa_1
B_{+}(\kappa;\kappa_1)
C(\kappa;\kappa_1)
\delta\left(\omega{-}D(\kappa;\kappa_1)\right)
\nonumber\\
&&\quad{+}
\int\limits_{-\pi}^\pi{\rm{d}}\kappa_1
B_{-}(\kappa;\kappa_1)
C(\kappa{+}\pi;\kappa_1)
\delta\left(\omega{-}D(\kappa{+}\pi;\kappa_1)\right),
\nonumber\\
&&B_{\pm}(\kappa;\kappa_1)=\left[g_\pm\left(u_{\kappa_1}u_{\kappa_1+\kappa} \pm v_{\kappa_1}v_{\kappa_1+\kappa}\right)
\right.
\nonumber\\
&&\left.
\quad \mp g_\mp\left(u_{\kappa_1}v_{\kappa_1+\kappa} \pm v_{\kappa_1}u_{\kappa_1+\kappa}\right) \right]^2,
\nonumber\\
&&C(\kappa;\kappa_1)=n_{\kappa_1}\left(1-n_{\kappa_1+\kappa}\right),
\nonumber\\
&&D(\kappa;\kappa_1)=\Lambda_{\kappa_1+\kappa}-\Lambda_{\kappa_1},
\end{eqnarray}
where $n_{\kappa}=1/\left(e^{\Lambda_{\kappa}/T}+1\right)$ is the Fermi-Dirac function for the spinless fermions (\ref{208}).
Hence, $S_{zz}(\kappa,\omega)$ is governed exclusively by two-fermion excitation continua.

\begin{figure}[htp]
\centering
\includegraphics[clip=on,width=\myfigsize]{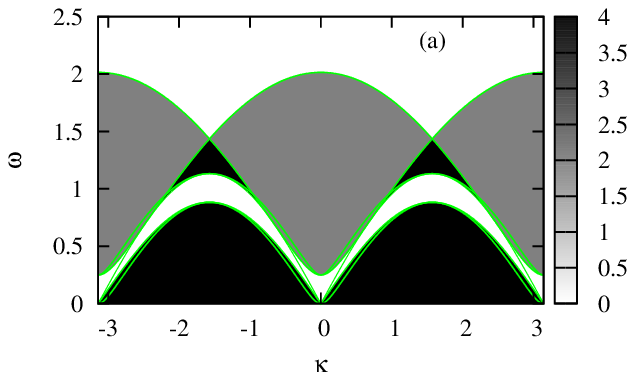}
\includegraphics[clip=on,width=\myfigsize]{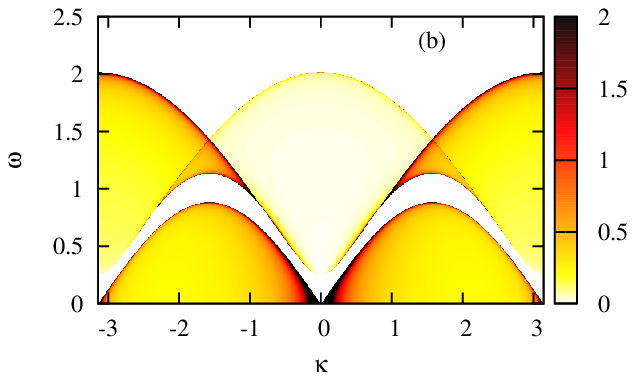}
\includegraphics[clip=on,width=\myfigsize]{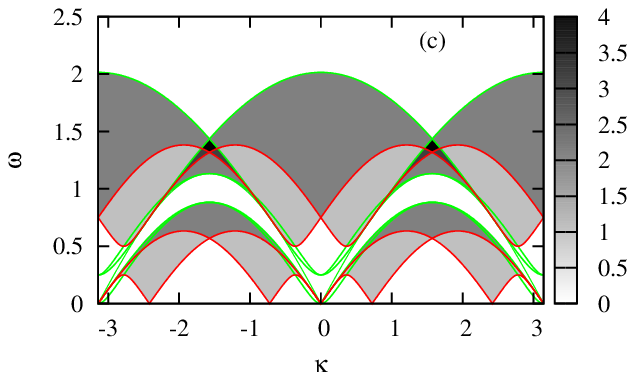}
\includegraphics[clip=on,width=\myfigsize]{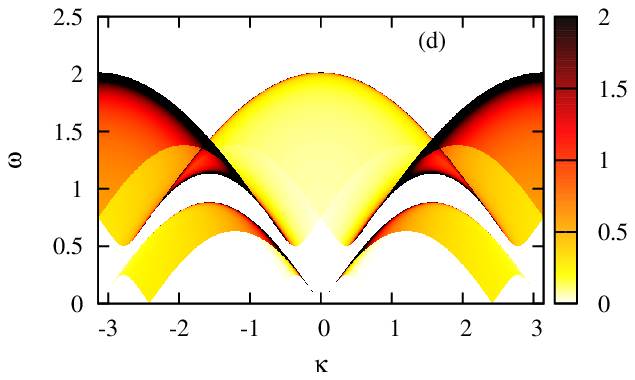}
\caption{(Color online)
Towards the dynamic structure factor $S_{zz}(\kappa,\omega)$.
$|J|=1$, $g_1=1$, $g_2=0.5$, $h=0.5$.
(a) Number of roots of two equations (\ref{506}).
(b) $S_{zz}(\kappa,\omega)$ at $T=\infty$.
(c) The same as in panel (a) but taking into accounting the Fermi-Dirac functions at $T=0$.
(d) $S_{zz}(\kappa,\omega)$ at $T=0$.
Green and red lines are the boundaries (\ref{a01}) and (\ref{a02}) correspondingly.}
\label{x05}
\end{figure}

Let us discuss this two-fermion quantity in more detail.
For fixed $\kappa$ and $\omega$, one has to solve the equations
\begin{eqnarray}
\label{506}
&&\omega-D(\kappa;\kappa_r)=0,
\;\;
\omega-D(\kappa+\pi;\kappa'_r)=0,
\end{eqnarray}
i.e., to find all roots $\kappa_r$, $\kappa'_r$.
Then Eq.~(\ref{505}) can be written down as follows:
\begin{eqnarray}
\label{507}
S_{zz}(\kappa,\omega)&{=}&
\sum_{\kappa_r}
\frac{B_{+}(\kappa;\kappa_r) C(\kappa;\kappa_r)}{A(\kappa;\kappa_r)}
\nonumber\\
&&{+}
\sum_{\kappa'_r}
\frac{B_{-}(\kappa;\kappa'_r) C(\kappa{+}\pi;\kappa'_r)}{A(\kappa{+}\pi;\kappa'_r)},
\end{eqnarray}
where
\begin{eqnarray}
\label{508}
&&A(\kappa;\kappa_1)
=
\left\vert
\frac{\partial D(\kappa;\kappa_1)}{\partial \kappa_1}
\right\vert
\\
&&{=}J^2
\left|\frac{|\cos(\kappa_1{+}\kappa)| \sin(\kappa_1{+}\kappa)}{\sqrt{J^2\cos^2(\kappa_1{+}\kappa){+}g_-^2h^2}}
{-}
\frac{|\cos\kappa_1| \sin\kappa_1}{\sqrt{J^2\cos^2\kappa_1{+}g_-^2h^2}}\right|.
\nonumber
\end{eqnarray}

\begin{figure}
\centering
\includegraphics[clip=on,width=42.5mm]{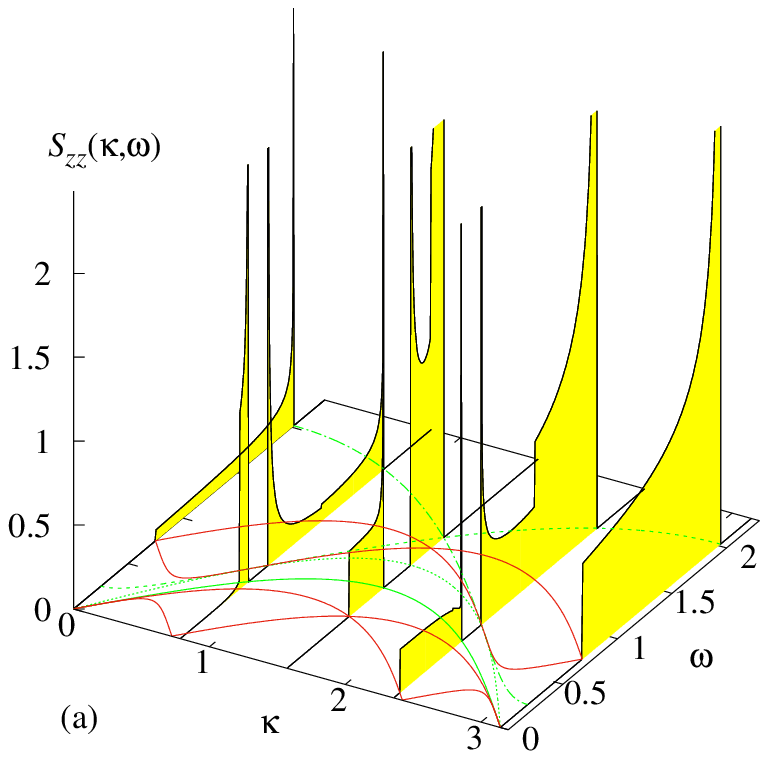}
\includegraphics[clip=on,width=42.5mm]{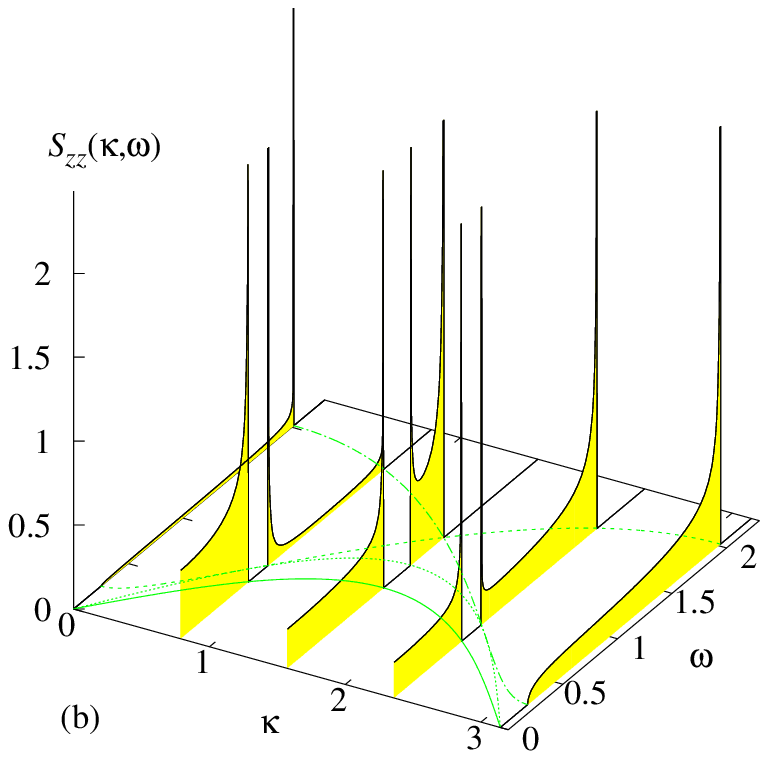}
\caption{(Color online)
$S_{zz}(\kappa,\omega)$ vs $\omega$
at
$\kappa=0$,
$\kappa=\pi/4$,
$\kappa=\pi/2$,
$\kappa=3\pi/4$,
and
$\kappa=\pi$.
$|J|=1$, $g_1=1$, $g_2=0.5$, $h=0.5$,
$T=0$ (left), cf. Fig.~\ref{x05}(d),
and
$T\to\infty$ (right), cf. Fig.~\ref{x05}(b).
Green and red lines are the boundaries (\ref{a01}) and (\ref{a02}) correspondingly.}
\label{x06}
\end{figure}

In Fig.~\ref{x05}(a) we show
(for a representative set of parameters)
the regions in the $\kappa$--$\omega$ plane
where equations (\ref{506}) have four roots (black), two roots (gray) or no roots (white).
In other words,
we plot $S_{zz}(\kappa,\omega)$ (\ref{507}) assuming
$A(\kappa;\kappa_1)=A(\kappa+\pi;\kappa_1)=1$ as well as $B_\pm(\kappa;\kappa_1)=1$ and $C(\kappa;\kappa_1)=C(\kappa+\pi;\kappa_1)=1$.
Clearly,
the dynamic structure factor $S_{zz}(\kappa,\omega)$ is identically zero within the white regions in the $\kappa$--$\omega$ plane
[equations (\ref{506}) have no roots].
Furthermore,
any two-fermion quantity have some structure coming from the factors $1/A(\kappa;\kappa_1)$ and $1/A(\kappa+\pi;\kappa_1)$.
It is nicely seen in the infinite-temperature limit when $C(\kappa;\kappa_1)=C(\kappa+\pi;\kappa_1)=1/4$ shown in Fig.~\ref{x05}(b).
Next,
deviating from the infinite-temperature limit
we have to examine the effect of the Fermi-Dirac functions in Eq.~(\ref{507})
which may suppress the dynamic structure factor $S_{zz}(\kappa,\omega)$ even in the gray or black regions,
especially at $T=0$.
In Fig.~\ref{x05}(c) we show the effect of the ground state Fermi-Dirac distributions for the same set of parameters
[we plot $S_{zz}(\kappa,\omega)$ (\ref{507}) assuming $A(\kappa;\kappa_1)=A(\kappa+\pi;\kappa_1)=1$ and $B_\pm(\kappa;\kappa_1)=1$].
In addition to the two- and four-roots regions,
the regions with one and three roots, surviving after the thermodynamic averaging, come into play
[compare Figs.~\ref{x05}(c) and \ref{x05}(a)].
Moreover, some allowed previously regions become white at $T=0$
signalizing the action of the Fermi-Dirac functions in the ground state.
The final gray-scale plot of the $zz$ dynamic structure factor (\ref{507}) at $T=0$ is presented in Fig.~\ref{x05}(d).
The frequency profiles for the chosen set of parameters are also plotted in Fig.~\ref{x06} complementing the gray-scale plot in Figs.~\ref{x05}(b,d).
It is clearly seen that the $zz$ dynamic structure factor at $T\to\infty$ shows the van Hove divergence at the edges of the two-fermion continua
which is typical for the $XX$ chains (see Refs.~\cite{Derzhko2000,Derzhko2002,Derzhko2008} for a review).
$S_{zz}(\kappa,\omega)$ in the ground state [Fig.~\ref{x05}(d)] demonstrates even richer behavior
due to the step-like form of the Fermi-Dirac functions [see Fig.~\ref{x05}(c,d) and Fig.~\ref{x06}(a)].
The analytical formulas for the boundaries of the two-fermion continua are given in Appendix.

\begin{figure}
\centering
\includegraphics[clip=on,width=\myfigsize]{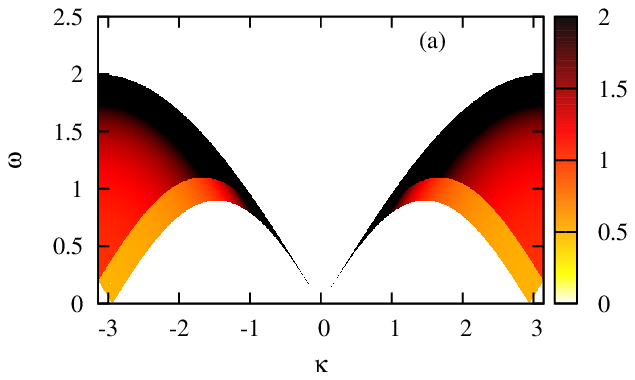}
\includegraphics[clip=on,width=\myfigsize]{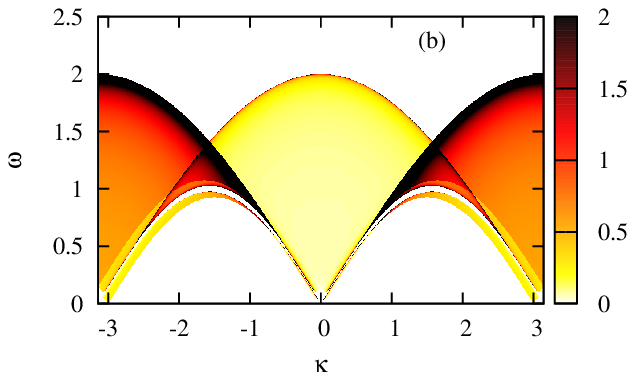}
\includegraphics[clip=on,width=\myfigsize]{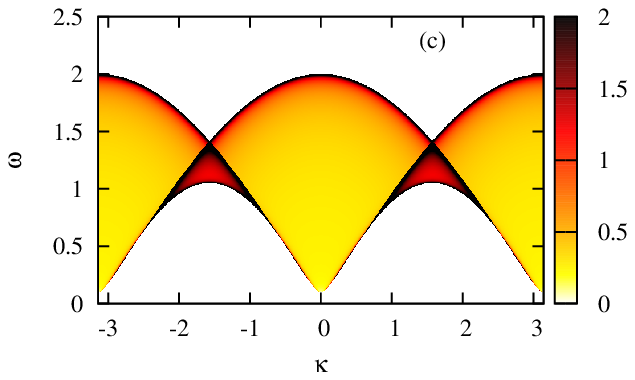}
\includegraphics[clip=on,width=\myfigsize]{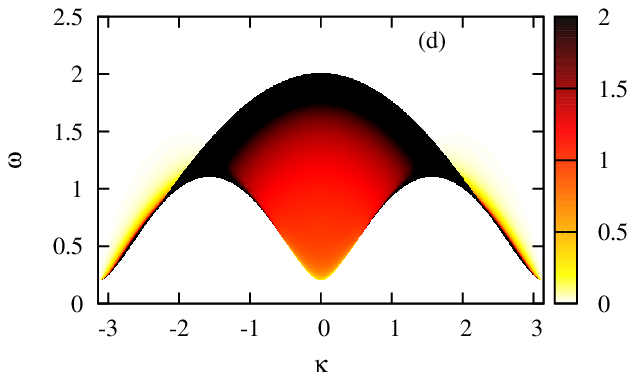}
\caption{(Color online)
The density plot of the dynamic structure factor $S_{zz}(\kappa,\omega)$ at $T=0$:
$|J|=1$, $g_1=1$,
$g_2=1$ (a),
$g_2=0.5$ (b),
$g_2=0$ (c),
$g_2=-1$ (d),
$h=0.1$.}
\label{x07}
\end{figure}

\begin{figure}
\centering
\includegraphics[clip=on,width=\myfigsize]{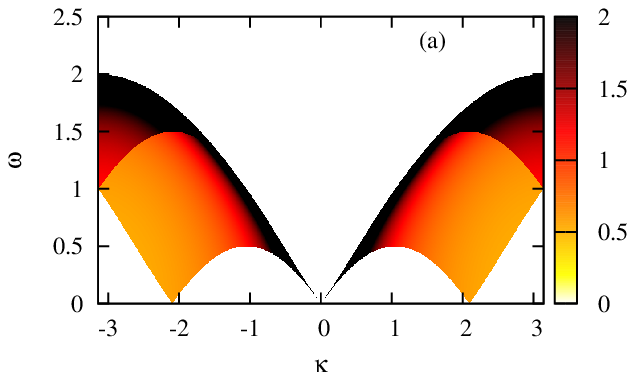}
\includegraphics[clip=on,width=\myfigsize]{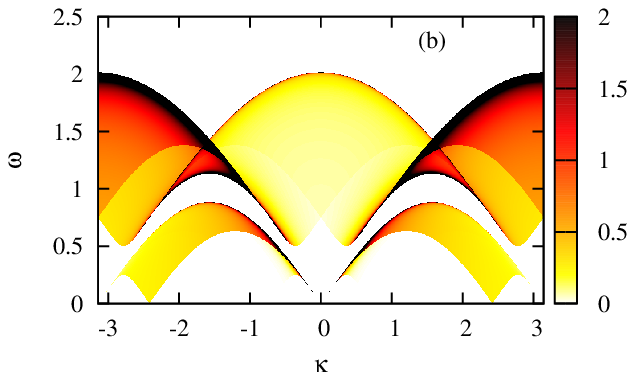}
\includegraphics[clip=on,width=\myfigsize]{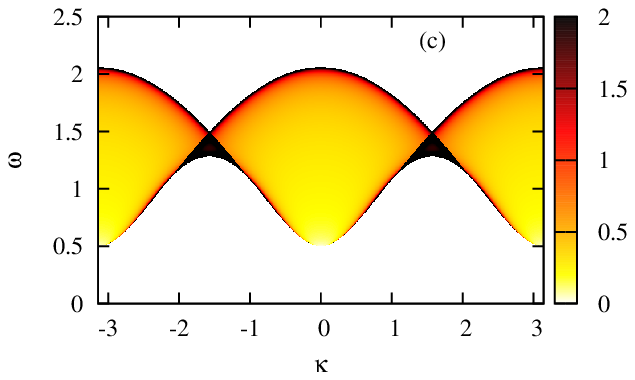}
\includegraphics[clip=on,width=\myfigsize]{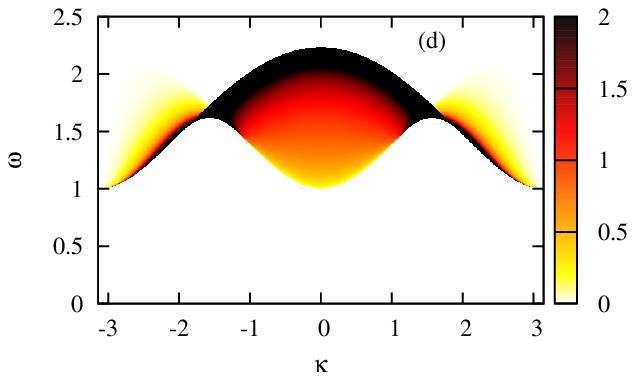}
\caption{(Color online)
The density plot of the dynamic structure factor $S_{zz}(\kappa,\omega)$ at $T=0$:
$|J|=1$, $g_1=1$,
$g_2=1$ (a),
$g_2=0.5$ (b),
$g_2=0$ (c),
$g_2=-1$ (d),
$h=0.5$.}
\label{x08}
\end{figure}

\begin{figure}
\centering
\includegraphics[clip=on,width=\myfigsize]{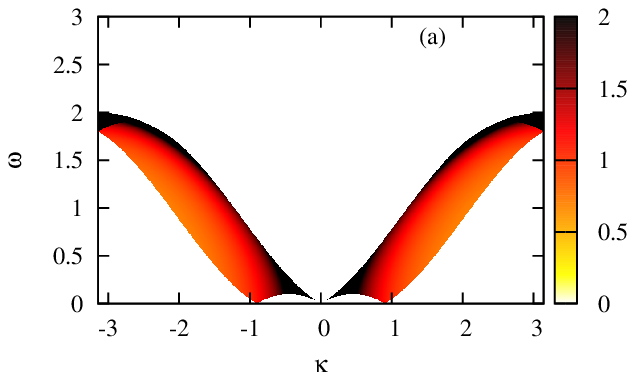}
\includegraphics[clip=on,width=\myfigsize]{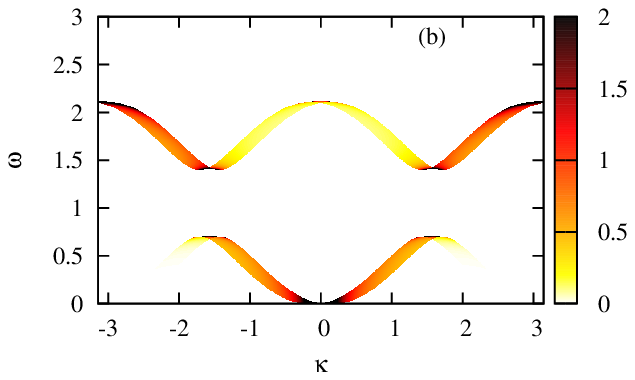}
\includegraphics[clip=on,width=\myfigsize]{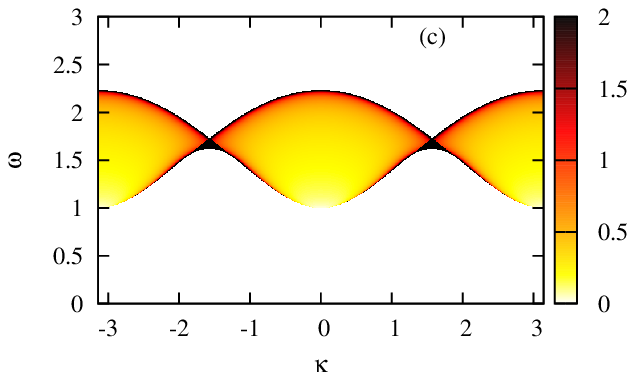}
\includegraphics[clip=on,width=\myfigsize]{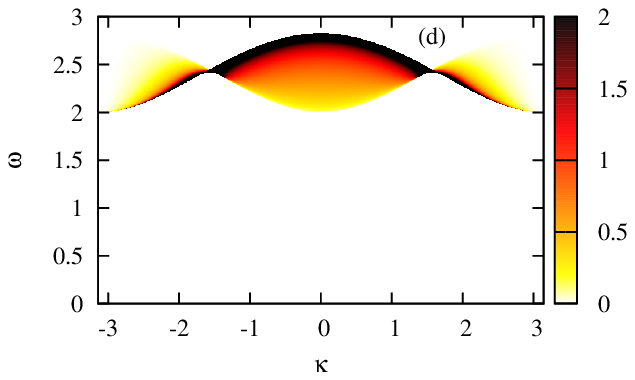}
\caption{(Color online)
The density plot of the dynamic structure factor $S_{zz}(\kappa,\omega)$ at $T=0$:
$|J|=1$, $g_1=1$,
$g_2=1$, $h=0.9$ (a),
$g_2=0.5$, $h=1.4$ (b),
$g_2=0$, $h=1$ (c),
$g_2=-1$, $h=1$ (d).}
\label{x09}
\end{figure}

We can understand the reported findings
taking into account that the dynamic structure factor $S_{zz}(\kappa,\omega)$ is governed by two-fermion continua.
The general effect of alternating $g$-factors can be understood from Figs.~\ref{x07}--\ref{x09},
where some results for $S_{zz}(\kappa,\omega)$ for different fields $h$ and values of $g_2$ at $T=0$ are collected.
The decreasing of $g_2$ from 1 to $-1$ at fixed value of magnetic field $h$ and $g_1=1$
leads to redistribution of the intensity of the $zz$ dynamic structure factor
from the boundary to the center of the Brillouin zone.
For $g_2 \in (0,1)$,
there are two regions with $S_{zz}(\kappa,\omega) \ne 0$ (top and bottom) which are disconnected,
see Figs.~\ref{x07}(b), \ref{x08}(b) and \ref{x09}(b).
The distances between these top and bottom regions increase with decreasing $g_2$ and with increasing $h$.
For $g_2 \in [-1,0]$, the increasing of the magnetic field $h$
leads to redistribution of the intensity of the $zz$ dynamic structure factor to higher frequencies.

Let us consider the effect of changes $g$-factors and $h$ in more detail.
At zero field,
the $zz$ structure factor is extremely simple
[see Eqs.~(\ref{502}) and (\ref{503})]
and can be presented as a sum of two contributions for the uniform model shifted by $\pi$ along the wave-vector axis
[i.e., Eq.~(\ref{502}) in the case of zero staggered spin structure factor $\overline{S}^{0}_{zz}(\kappa,\omega)$].
It is definitely also the case of a small field (see Fig.~\ref{x07} for $h=0.1$).
It is clearly seen that at small $h$, the deviation of $g_2$ from $g_1=1$
induces a tiny strip of new two-fermion continuum at lower frequencies.
The intensity of this low-energy two-fermion continuum wanes with decreasing $g_2$.
Surprisingly, $S_{zz}(\kappa,\omega)$ for $g_2\leq0$ does not show any trace of the low-energy continuum anymore
[see Figs.~\ref{x07}(c,d)]: The $zz$ structure factor shows one two-fermion continuum only.
In contrast to $h=0$,
at small fields, two opposite cases $g_2=1$ and $g_2=-1$ are not identical
[compare Fig.~\ref{x07}(a) and Fig.~\ref{x07}(d)].

At higher fields,
the magnetic structure factor cannot be approximated by the sum of uniform spin structure factors ${S}^{0}_{zz}(\kappa,\omega)$ anymore.
Even for a moderate alternation of $g$-factors [$g_1=1$, $g_2=0.5$ in Fig.~\ref{x08}(b)]
we observe the appearance of another two-fermion continuum at lower frequencies.
It can be treated as a splitting of the initial continuum inherent in the uniform model [see Fig.~\ref{x08}(a)] in two parts,
which is a signal of the two-band structure of the fermion excitation spectrum (\ref{208}).
It should be noted that the two-fermion continuum at lower frequencies induced by small deviation of $g_2$ (from $g_1=1$)
is not a tiny strip anymore as it was at small fields ($h=0.1$).
At higher fields as well as at small ones,
the $zz$ structure factor for $g_2\leq0$ shows just one two-fermion continuum only [Figs.~\ref{x08}(c,d)].
This picture keeps the tendency with increasing field as it is shown in Fig.~\ref{x09}.
In two top panels we present results at magnetic fields close to $h_s$ whereas for $g_1g_2\leq 0$ we put $h=1$ [Figs.~\ref{x09}(c,d)],
because at $g_1g_2\leq 0$ the saturation field does not exist.
The fact, that in Fig.~\ref{x09}(b) both the low-energy and hight-energy two-fermion continua are tiny strips,
is caused by that the field is very close to $h_s$.

\begin{figure}
\centering
\includegraphics[clip=on,width=\myfigsize]{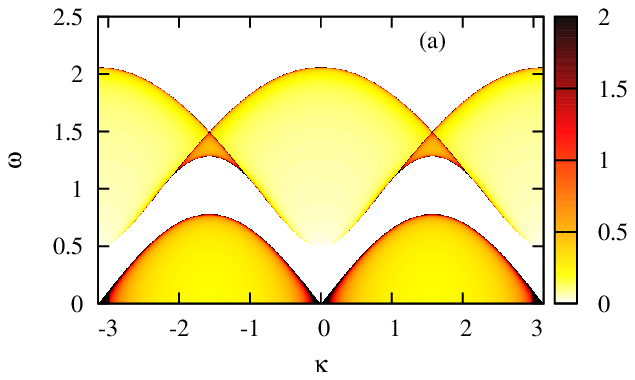}
\includegraphics[clip=on,width=\myfigsize]{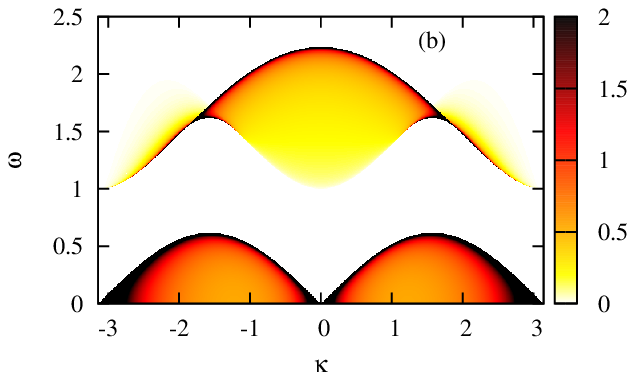}
\caption{(Color online)
The density plot of the dynamic structure factor $S_{zz}(\kappa,\omega)$ at $T\to\infty$:
$|J|=1$, $g_1=1$, $h=0.5$,
$g_2=0$ (a),
$g_2=-1$ (b).}
\label{x10}
\end{figure}

We also examine the temperature effect on the $zz$ structure factor for non-positive $g_2\le 0$.
The results for $T\to\infty$ in Fig.~\ref{x10} show an additional two-fermion continuum for low frequencies.
In case of zero temperature this continuum was hidden owing to the Fermi-Dirac functions,
compare Fig.~\ref{x10} to Fig.~\ref{x08}.

In the case $\kappa=0$, Eq.~(\ref{505}) can be transformed to the following form:
\begin{eqnarray}
\label{509}
&&S_{zz}(0,\omega){=}
\delta(\omega)\int\limits_{-\pi}^{\pi}{\rm d}\kappa_1(g_{+}^{} {-} 2g_{-}^{}u_{\kappa_1}v_{\kappa_1})^2
n_{\kappa_1}(1{-}n_{\kappa_1})
\nonumber\\
&&+\frac{g_{-}^2\sqrt{\omega^2 {-} 4g_-^2h^2}}{\omega\sqrt{4J^2 {+} 4g_-^2h^2 {-} \omega^2}}
\sum_{\kappa_r}n_{\kappa_r}\left(1 {-} n_{\kappa_r+\pi}\right),
\end{eqnarray}
where $\kappa_r$ are solutions of the equation $\omega=\Lambda_{\kappa_r+\pi}-\Lambda_{\kappa_r}$.
The latter equation has solutions only in the restricted region
\begin{eqnarray}
\label{510}
2\vert g_-h\vert\le\omega <2\sqrt{J^2+g_-^2h^2}.
\end{eqnarray}

We can use Eqs.~(\ref{509}) and (\ref{504}) to get explicit expressions for the absorption intensity $I_z(\omega,h)$:
\begin{eqnarray}
\label{511}
&&I_{z}(\omega,h) \propto
\frac{g_{-}^2\sqrt{\omega^2-4g_-^2h^2}}{\sqrt{4J^2+4g_-^2h^2-\omega^2}}
\\
&&\times
\frac{1-\exp(-\beta\omega)}{(1+\exp[\beta\left(g^{}_+h-\frac{\omega}{2}\right)])
(1+\exp[-\beta\left(g^{}_+h+\frac{\omega}{2}\right)])}.
\nonumber
\end{eqnarray}
In the ground state we arrive at the following formula:
\begin{eqnarray}
\label{512}
&&I_{z}(\omega,h) \propto
\frac{g_{-}^2\sqrt{\omega^2-4g_-^2h^2}}{\sqrt{4J^2+4g_-^2h^2-\omega^2}},
\end{eqnarray}
where in case $g_1g_2>0$ the Fermi-Dirac functions shrink further the condition of allowed $\omega$
[see Eq.~(\ref{510})]
to the following one: $2\vert g_+h\vert<\omega< 2\sqrt{J^2+g_-^2h^2}$.

\begin{figure}
\begin{center}
\includegraphics[clip=on,width=0.49\columnwidth]{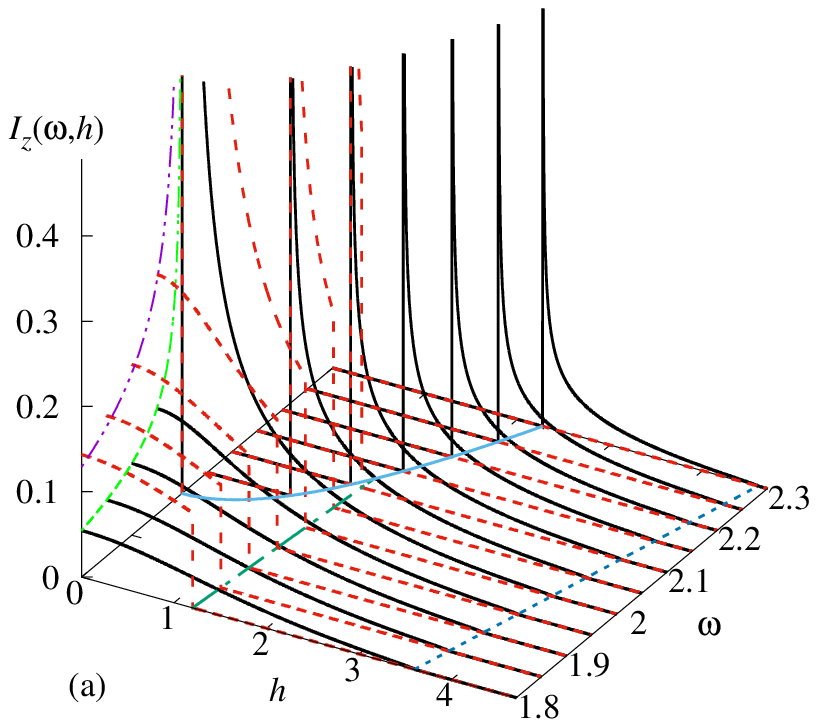}
\includegraphics[clip=on,width=0.49\columnwidth]{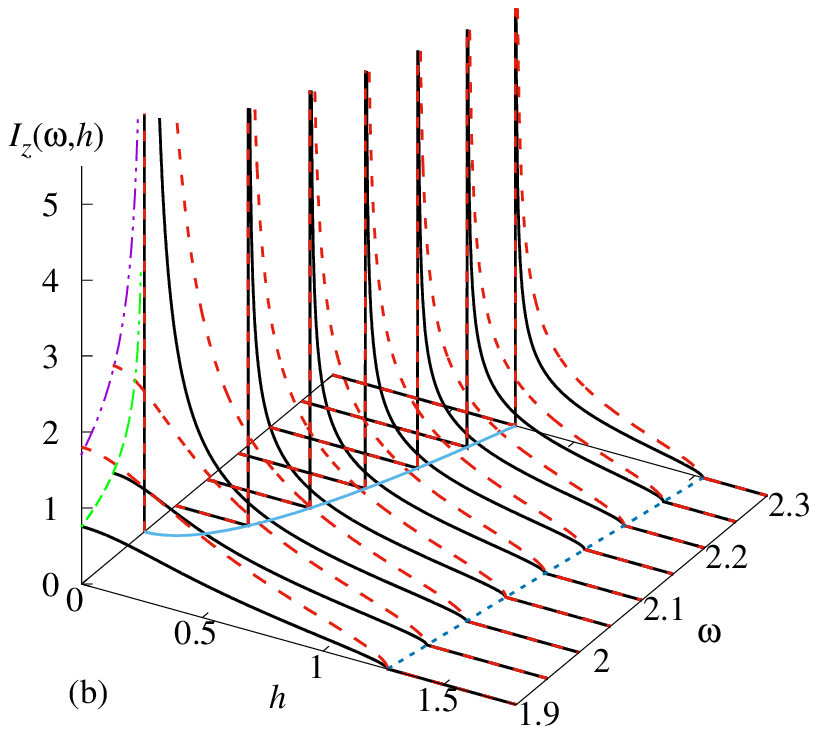}
\caption
{(Color online)
Field profiles of the absorption intensity $I_z(\omega,h)$ at different frequencies $\omega$ for $|J|=1$, $g_1=1$,
$g_2=0.5$ (a),
$g_2=-0.5$ (b),
and temperatures $T=1$ (solid black curves) and $T=0$ (dashed red curves).
The dashed-dot-dot violet (dashed green) curve indicates the intensity at $h=0$ and $T=0$ ($T=1$).
The solid and short-dashed blue curves show the boundaries given in Eq.~(\ref{510})
while the dashed-dot green curve in panel (a), given by $h=\omega/(2|g_{+}|)$, denotes the upper boundary of $I_z(\omega,h)$ at $T=0$
(see the discussion in the text).}
\label{x11}
\end{center}
\end{figure}

It is evident from Eq.~(\ref{511}) that there is no energy absorption in case of the uniform $g$-factors ($g_1=g_2=1$),
since the total magnetization commutes with the Hamiltonian.
The alternation of $g$-factors destroys this property and leads immediately to nonzero absorption intensity $I_z(\omega,h)$.
From Eqs.~(\ref{512}) and (\ref{511}) one can deduce the shape of the absorption line.
The field profiles of the absorption intensity for alternating $g$-factors are shown in Fig.~\ref{x11}.
The absorption intensity curve $I_z(\omega,h)$ for any frequency ends continuously at $h=\omega/(2|g_{-}|)$
for both $T=0$, $g_1g_2<0$ and $T>0$ cases.
It is clearly seen in Figs.~\ref{x11}(a,b); short-dashed blue line.
If the frequency exceeds $2|J|$,
we observe also a van Hove singularity at $h=\sqrt{\omega^2-4J^2}/(2|g_{-}|)$
[see Figs.~\ref{x11}(a,b); solid blue line].
In the ground state for $g_1g_2>0$ this singularity disappears at $\omega=2|J|/\sqrt{1-(g_{-}/g_{+})^2}$.
If $\omega<2|J|/\sqrt{1-(g_{-}/g_{+})^2}$ for zero temperature and $g_1g_2>0$,
the absorption intensity curve $I_z(\omega,h)$ ends abruptly at $h=\omega/(2|g_{+}|)$
[see Fig.~\ref{x11}(a); dashed-dot green line],
and at $\omega>2|J|/\sqrt{1-(g_{-}/g_{+})^2}$ this ground-state absorption intensity vanishes, $I_z(\omega,h)=0$.

\subsection{$xx$ dynamics}

We pass to another dynamic structure factor, namely, the $xx$ structure factor $S_{xx}(\kappa,\omega)$.
We perform the computation of the $xx$ time correlation functions numerically
using the previously elaborated method \cite{Derzhko1997,Derzhko2000,Derzhko2008}.
In what follows, we consider the finite chain of $N=400$ spins with open boundary conditions.
To avoid the boundary effect, we have to adapt Eq.~(\ref{501}).
Thus, we choose a ``central'' spin at the site $j=61,81$ (depending on the adopted parameters)
and then calculate the time correlation functions $\langle s^x_j(t)s^x_{j+n}\rangle$ as well as $\langle s^x_{j+1}(t)s^x_{j+n+1}\rangle$ for $n\geq 0$.
Finally, we present the Fourier transform in Eq.~(\ref{501}) in the following symmetrized form:
\begin{eqnarray}
\label{513}
&&S_{xx}(\kappa,\omega)
=
\frac{1}{2}\text{Re}\int_0^{\infty}{\rm d}t e^{-\epsilon t}e^{i\omega t}
\\
&&\times
\left\{
g_1^2\left[\langle s_j^x(t)s_j^x\rangle +2\sum_{n=1}^{\frac{N}{2}}\cos(2n\kappa)\langle s^x_j(t)s^x_{j+2n}\rangle \right]
\right.
\nonumber\\
&&+ 2g_1g_2\sum_{n=1}^{\frac{N}{2}}\cos((2n-1)\kappa)\langle s^x_j(t)s^x_{j+2n-1}\rangle
+
\nonumber\\
&&g_2^2\left[\langle s_{j+1}^x(t)s^x_{j+1}\rangle {+}2\sum_{n=1}^{\frac{N}{2}}\cos(2n\kappa)\langle s^x_{j+1}(t)s^x_{j+1+2n}\rangle \right]
\nonumber\\
&&\left.
{+} 2g_1g_2\sum_{n=1}^{\frac{N}{2}}\cos((2n{-}1)\kappa)\langle s^x_{j+1}(t)s^x_{j+1+2n-1}\rangle
\right\}.\nonumber
\end{eqnarray}
In numerical calculations we restrict the sum over $n$ up to $10 \ldots 50$ depending on the correlation length.

\begin{figure}
\centering
\includegraphics[clip=on,width=\myfigsize]{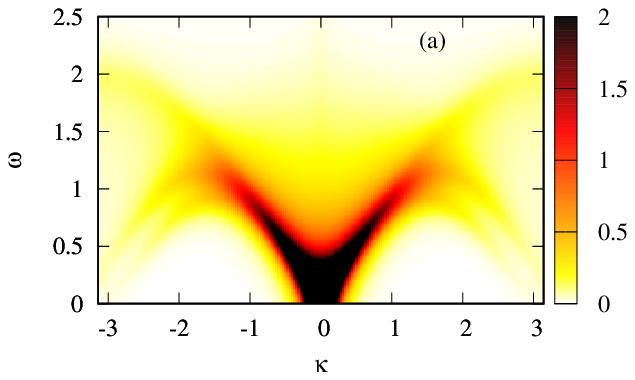}
\includegraphics[clip=on,width=\myfigsize]{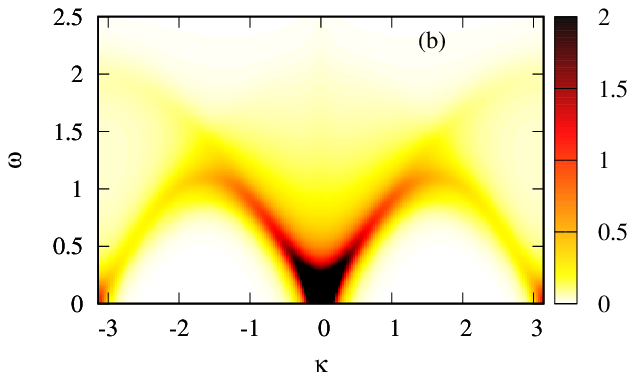}
\includegraphics[clip=on,width=\myfigsize]{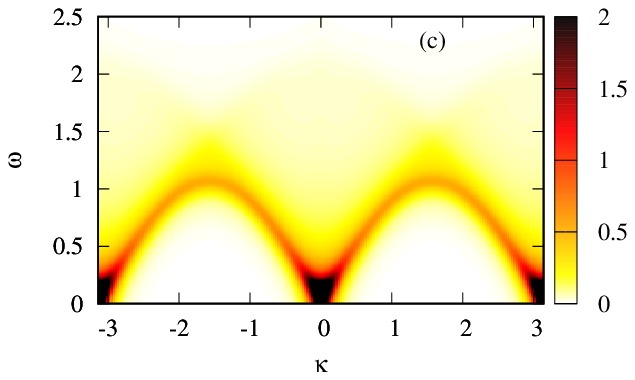}
\includegraphics[clip=on,width=\myfigsize]{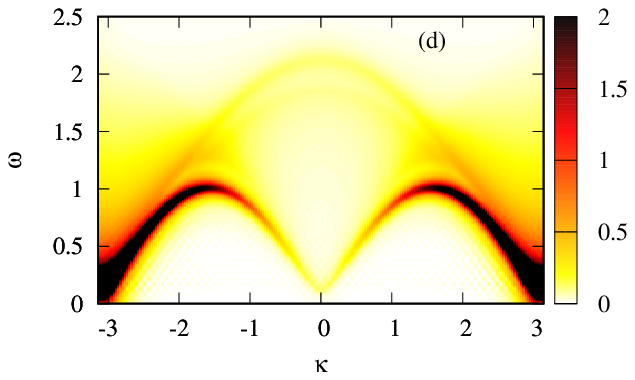}
\caption{(Color online) The density plot of the dynamic structure factor $S_{xx}(\kappa,\omega)$.
$J=-1$, $g_1=1$,
$g_2=1$ (a),
$g_2=0.5$ (b),
$g_2=0$ (c),
$g_2=-1$ (d),
$h=0.1$ at low temperature $T=0.1$.}
\label{x12}
\end{figure}

\begin{figure}
\centering
\includegraphics[clip=on,width=\myfigsize]{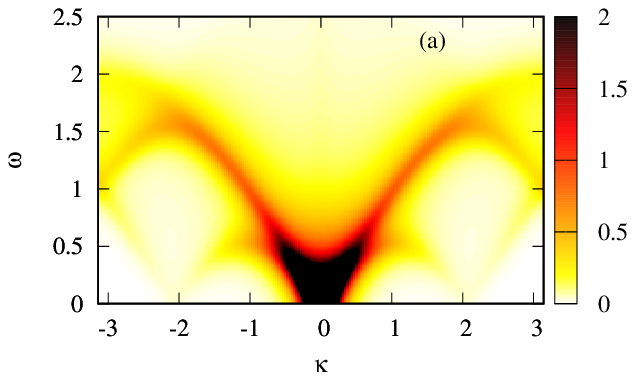}
\includegraphics[clip=on,width=\myfigsize]{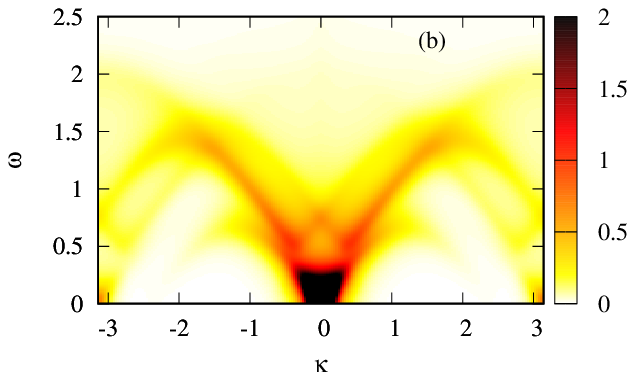}
\includegraphics[clip=on,width=\myfigsize]{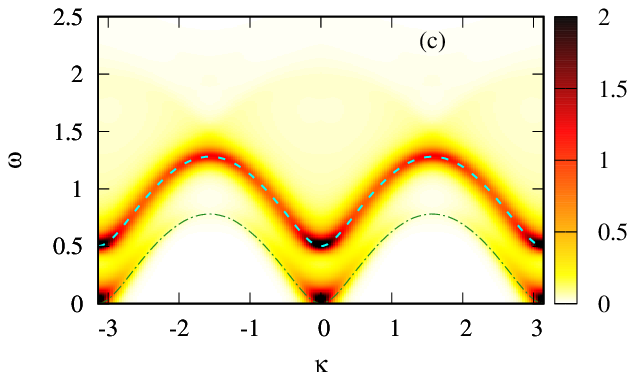}
\includegraphics[clip=on,width=\myfigsize]{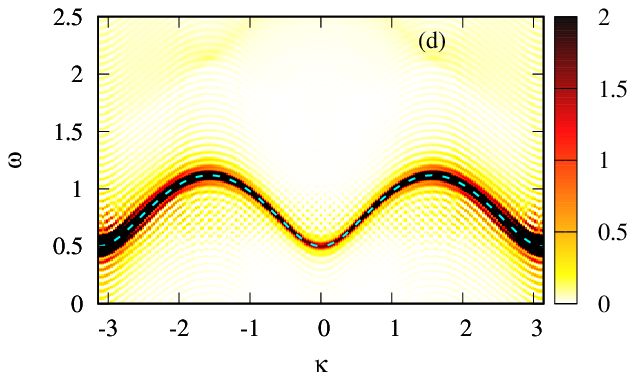}
\caption{(Color online) The density plot of the dynamic structure factor $S_{xx}(\kappa,\omega)$.
$J=-1$, $g_1=1$,
$g_2=1$ (a),
$g_2=0.5$ (b),
$g_2=0$ (c),
$g_2=-1$ (d),
$h=0.5$ at low temperature $T=0.1$.
Dashed and dashed-dot curves follow Eq.~(\ref{514}).}
\label{x13}
\end{figure}

\begin{figure}
\centering
\includegraphics[clip=on,width=\myfigsize]{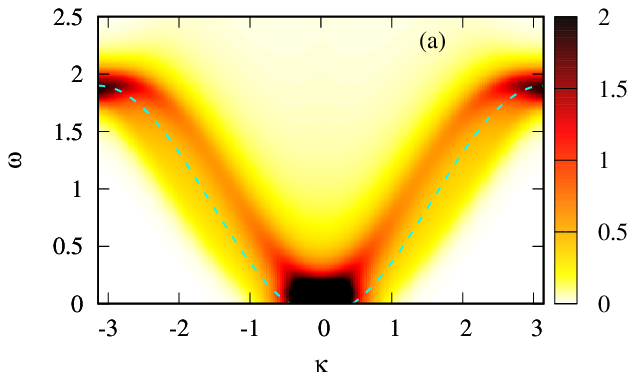}
\includegraphics[clip=on,width=\myfigsize]{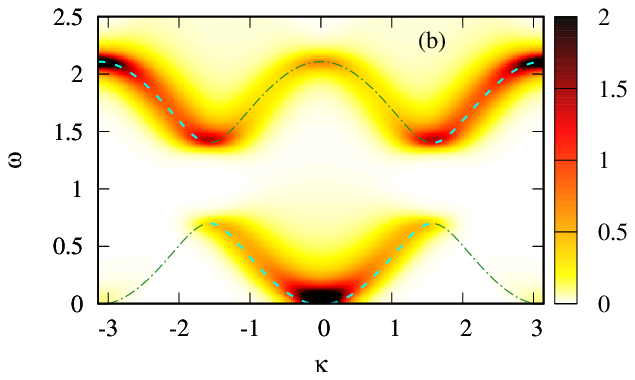}
\includegraphics[clip=on,width=\myfigsize]{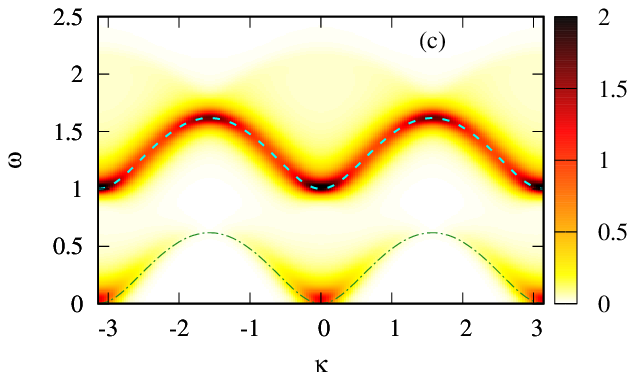}
\includegraphics[clip=on,width=\myfigsize]{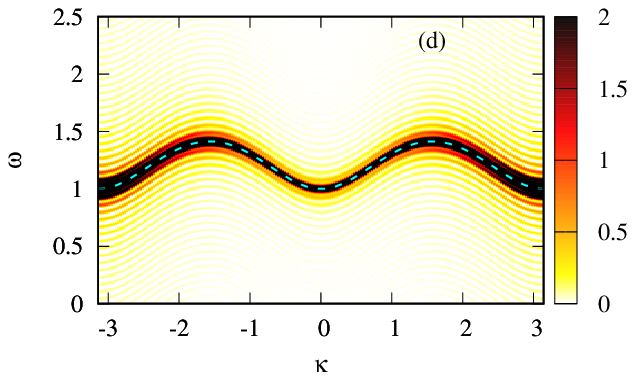}
\caption{(Color online) The density plot of the dynamic structure factor $S_{xx}(\kappa,\omega)$.
$J=-1$, $g_1=1$,
$g_2=1$, $h=0.9$ (a),
$g_2=0.5$, $h=1.4$ (b),
$g_2=0$, $h=1$ (c),
$g_2=-1$, $h=1$ (d) at low temperature $T=0.1$.
Dashed and dashed-dot curves in panels (a) and (b) correspond to $-\Lambda_{\kappa+\pi}$ and $-\Lambda_\kappa$.
Dashed and dashed-dot curves in panels (c), (d) follow Eq.~(\ref{514}).}
\label{x14}
\end{figure}

\begin{figure}
\centering
\includegraphics[clip=on,width=42.5mm]{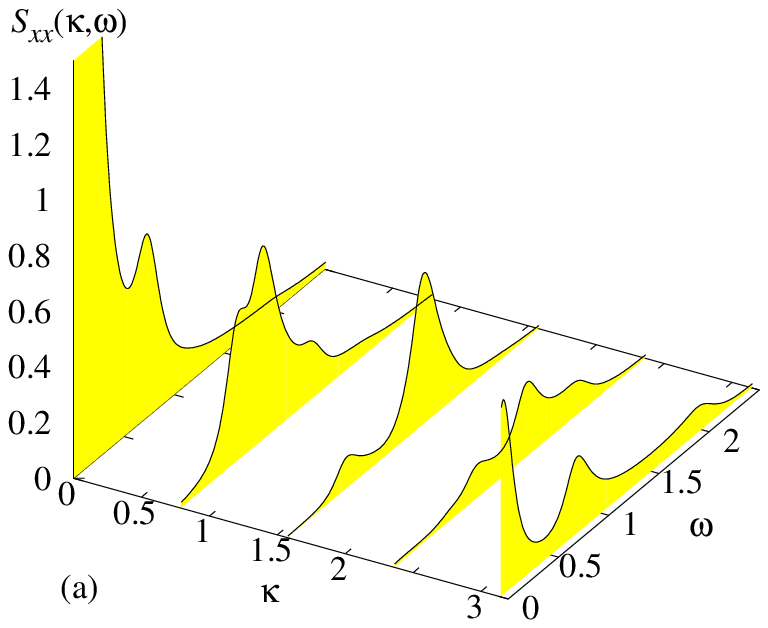}
\includegraphics[clip=on,width=42.5mm]{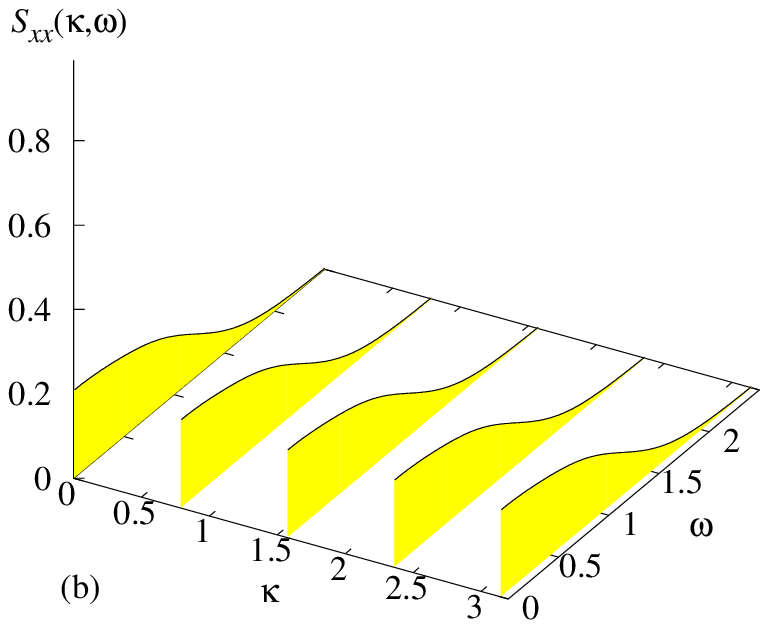}
\caption{(Color online) $S_{xx}(\kappa,\omega)$ vs $\omega$
at
$\kappa=0$,
$\kappa=\pi/4$,
$\kappa=\pi/2$,
$\kappa=3\pi/4$,
and
$\kappa=\pi$.
$J=-1$, $g_1=1$, $g_2=0.5$, $h=0.5$,
$T=0.1$ (left panel), cf. Fig.~\ref{x13},
and
$T\to\infty$ (right panel).}
\label{x15}
\end{figure}

The results of the numerical calculation for $S_{xx}(\kappa,\omega)$ at sufficiently low temperature $T=0.1$
are shown in Figs.~\ref{x12}--\ref{x15}.
In contrast to the $zz$ structure factor,
$S_{xx}(\kappa,\omega)$ is not governed exclusively by the continuum of two-fermion excitations.
However,
the deeper inspection of Figs.~\ref{x12}--\ref{x15} reveals some resemblance between the $zz$ and $xx$ structure factors.
Although there is no singular parts visible in $S_{xx}(\kappa,\omega)$ as well as abrupt boundaries for the regions with nonzero values,
the dominating contribution in the case of positive $g_2$ is circumscribed by the boundaries of the two-fermion continua outlined in Appendix.
The same feature was demonstrated earlier for the uniform and dimerized $XX$ chains \cite{Derzhko2000,Derzhko2002}.
We can deduce from relation (\ref{502}) and Fig.~\ref{x12} that the staggered spin structure factor (\ref{503}) is minor at small fields.
Thus,
one can observe how the intensity of the structure factor $S_{xx}(\kappa,\omega)$ is redistributed between two basic continua of the uniform chain
[see Fig.~\ref{x12}(a)]
shifted by $\pi$ with respect to each other
when $g_2$ decreases from 1 up to negative values.
One can still recognize the similar feature even at intermediate field $h=0.5$ in case of $g_2>0$ in Fig.~\ref{x13}(b)
where the combination of two continua of $S^0_{xx}(\kappa,\omega)$ and $S^0_{xx}(\kappa+\pi,\omega)$ creates an intricate intensity picture.

Interestingly, the structure factor $S_{xx}(\kappa,\omega)$ for non-positive $g_2\leq 0$ is concentrated mainly along the lines
\begin{eqnarray}
\label{514}
\lambda^{\pm}_{\kappa}=\sqrt{J^2\sin^2\kappa+g_-^2h^2}\pm g_+h.
\end{eqnarray}
Although the exact $xx$ correlation functions and the exact $xx$ structure factor are not known for $g_1g_2<0$,
one can adapt the procedure of Refs.~\cite{cruz1981,Derzhko2002} for the case of the uniform and dimerized chains above the saturation field.
We need to make the crucial assumption
that the action of the Jordan-Wigner phase factors on the ground state is equivalent to its action on the ideal antiferromagnetic state.
Then, the problem is reduced to calculation of the pair correlation functions for spinless fermions
with the final result
\begin{eqnarray}
\label{515}
&&\!\!\!\!\!S_{xx}(\kappa,\omega){\approx}
\\
&&\frac{\pi}{4}\left\{\!
\left(g_+^2 {+} g_-^2 {+} 4g^{}_+g^{}_-{\rm sgn}(h)u_{\kappa{+}\pi/2}|v_{\kappa{+}\pi/2}|\right)\!\delta(\omega{-}\lambda_{\kappa}^+)
\right.
\nonumber\\
&&\left.
{+}\left(g_+^2 {+} g_-^2 {-} 4g^{}_+g^{}_-{\rm sgn}(h)u_{\kappa+\pi/2}|v_{\kappa+\pi/2}|\right)\!\delta(\omega{-}\lambda_{\kappa}^-)
\!\right\}.
\nonumber
\end{eqnarray}
Equation~(\ref{515}) although approximate,
agrees with numerics shown in Figs.~\ref{x13}, \ref{x14} for negative $g_2$ (dashed and dashed-dot lines).

If $g_2 \in (0,1]$ for magnetic fields close to $h_s$,
the many-fermion continua shrink [see Fig.~\ref{x14}(a,b)]
and above the saturation fields they reduce to the one-fermion excitation spectrum
shifted by $\pi$ along the $\kappa$ axis with the reversed sign
[i.e., $-\Lambda_{\kappa+\pi}$, dashed line in Fig.~\ref{x14}(a,b)]
and if $g_2 \in (0,1)$,
also by the one-fermion excitation spectrum multiplied by $-1$
[i.e., $-\Lambda_{\kappa}$, dashed-dot line in Fig.~\ref{x14}(b)],
\begin{eqnarray}
\label{516}
S_{xx}(\kappa,\omega)&{=}&\frac{\pi}{2}\left[(g_{+}u_\kappa {-} g_{-}v_\kappa)^2\delta(\omega{-}\Lambda_\kappa)
\right.\\ && \left.
{+} (g_{+}v_\kappa {+} g_{-}u_\kappa)^2\delta(\omega{-}\Lambda_{\kappa{+}\pi}) \right],\: {\rm if}\: h<-h_s,
\nonumber\\
S_{xx}(\kappa,\omega)&{=}&\frac{\pi}{2}\left[(g_{+}v_\kappa {-} g_{-}u_\kappa)^2\delta(\omega{+}\Lambda_{\kappa})
\right.\nonumber\\ && \left.
{+} (g_{+}u_\kappa {+} g_{-}v_\kappa)^2\delta(\omega{+}\Lambda_{\kappa{+}\pi})
\right],\: {\rm if}\: h>h_s.
\nonumber
\end{eqnarray}
In case of $g_2\leq 0$,
in Fig.~\ref{x14}(c,d) we observe for higher field even more pronounced mode along the lines given in Eq.~(\ref{514}).

In Fig.~\ref{x15} we show the frequency profiles of the structure factor for several values of $\kappa=0,\pi/4,\pi/2,3\pi/4,\pi$.
It is clearly seen there
that the non-uniform $g$-factor leads to many-peak structure in the frequency dependences of $S_{xx}(\kappa,\omega)$ at the low temperature $T=0.1$,
see Fig.~\ref{x15}(a).
In contrast,
the infinite temperature
smears out the fine structure of $S_{xx}(\kappa,\omega)$ transforming the frequency profiles into $\kappa$-independent Gaussian ridges,
see Fig.~\ref{x15}(b).
Such a form can be obtained using the exact results for the time correlation functions of dimerized chain \cite{Perk1980}.
Those correlation functions vanish if the sites are different that leads to a $\kappa$-independent structure factor $S_{xx}(\kappa,\omega)$.
Utilizing the result of Ref.~\cite{Perk1980},
we get the following explicit formula for $S_{xx}(\kappa,\omega)$ at $T\to\infty$:
\begin{eqnarray}
\label{517}
&&S_{xx}(\kappa,\omega)
=
\frac{1}{8}\int\limits_{-\infty}^{\infty}{\rm{d}}t
e^{i\omega t}
\text{Re}\left\{g_1^2Z_o(t)+g_2^2Z_e(t)\right\},
\\
&&Z_e(t)
{=}
\frac{\theta_3(z,q)}{\theta_3(z_0,q)}
\frac{\theta_2(z^\prime,q)}{\theta_2(z^\prime_0,q)}
{\exp}\!{\left[ i g_{+}^{} ht{-}{\left(\!1{-}\frac{{\rm{E}}(\widetilde{\varkappa})}{{\rm{K}}(\widetilde{\varkappa})}\right)}J_{+}^2t^2\right]},
\nonumber\\
&&Z_o(t)=\exp\left(i2g_{+}ht\right)Z^{*}_e(t),
\nonumber\\
&&J_{\pm}=\frac{1}{2}\left(\sqrt{J^2+g_{-}^2h^2} \pm |g_{-}h| \right),
\nonumber\\
&&\widetilde{\varkappa}=\frac{J_-}{J_+}=\frac{J^2}{\left(\sqrt{J^2+g_{-}^2h^2} + |g_{-}h| \right)^2},
\nonumber\\
&&q=\exp\left(-\frac{\pi{\rm{K}}(\sqrt{1-\widetilde{\varkappa}^2})}{{\rm{K}}(\widetilde{\varkappa})}\right),\nonumber
\end{eqnarray}
where $\theta_2(z^\prime,q)$, $\theta_3(z,q)$ are the Jacobi theta-functions (see \cite{Perk1980} and references therein)  with
\begin{eqnarray}
\label{518}
&&z=\frac{\pi(J_+t+iv_0)}{2{\rm{K}}(\widetilde{\varkappa})},\; z^\prime=\frac{\pi(J_+t-iv_0)}{2{\rm{K}}(\widetilde{\varkappa})}
\nonumber\\
&&z_0=\frac{i\pi v_0}{2{\rm{K}}(\widetilde{\varkappa})},\; z_0^\prime=-\frac{i\pi v_0}{2{\rm{K}}(\widetilde{\varkappa})},
\end{eqnarray}
and the parameter $v_0$ is defined by the following relation:
\begin{eqnarray}
\label{519}
{\rm dc}(i v_0,\widetilde{\varkappa})=\frac{J}{2J_+},
\end{eqnarray}
where ${\rm dc}(i v_0,\widetilde{\varkappa})={\rm dn}(v_0, 1-\widetilde{\varkappa}^2)$ is the elliptic delta amplitude function for imaginary argument.

In case of strong magnetic field $h$ and non-uniform $g$-factors $g^{}_{-}\neq 0$ we have $\tilde\varkappa\ll 1$.
Expanding the correlation functions for small $\tilde\varkappa$,
we get the $xx$ structure factor in the explicit Gaussian form:
\begin{eqnarray}
\label{520}
S_{xx}(\kappa,\omega)&\approx& \frac{\sqrt{2\pi}}{4|J|}
\left[A_{-}\left(e^{-\frac{(\omega+\omega_{-})^2}{2J_{-}^2}} + e^{-\frac{(\omega-\omega_{-})^2}{2J_{-}^2}}\right)
\right.
\nonumber\\
&&\left.
+A_{+}\left(e^{-\frac{(\omega+\omega_{+})^2}{2J_{-}^2}} + e^{-\frac{(\omega-\omega_{+})^2}{2J_{-}^2}}\right)
\right],
\nonumber\\
\omega_{\pm}&=&J_{+}\pm g_{+}h,
\nonumber\\
A_{\pm}&=&(g_{+}^2+g_{-}^2)\frac{J_{+}}{|J|} \pm g_{+}g_{-}\sqrt{\frac{4J_{+}^2}{J^2}-1}.
\end{eqnarray}
From Eq.~(\ref{520}) it is clear that the intensity of the $xx$ structure factor in the infinite-temperature limit
is concentrated near two Gaussian peaks at $\omega=\omega_{\pm}$.

\begin{figure}
\begin{center}
\includegraphics[clip=on,width=42.5mm]{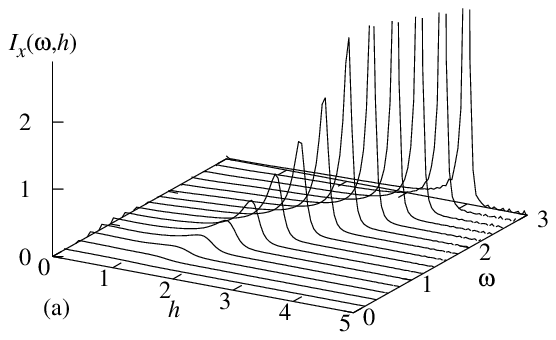}
\includegraphics[clip=on,width=42.5mm]{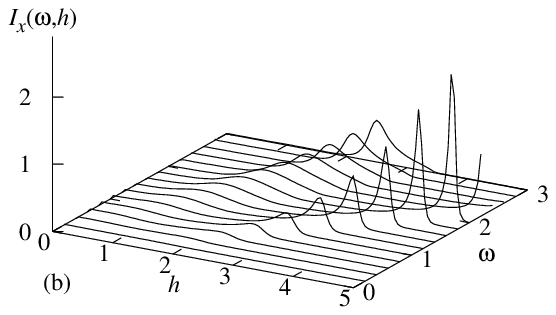}
\includegraphics[clip=on,width=42.5mm]{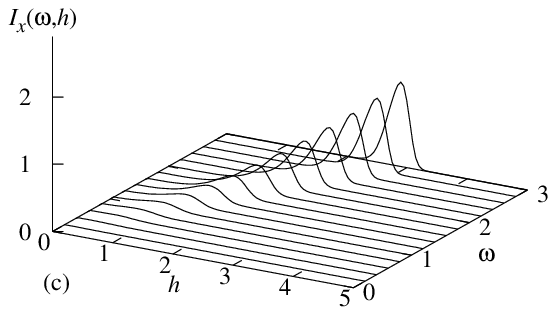}
\includegraphics[clip=on,width=42.5mm]{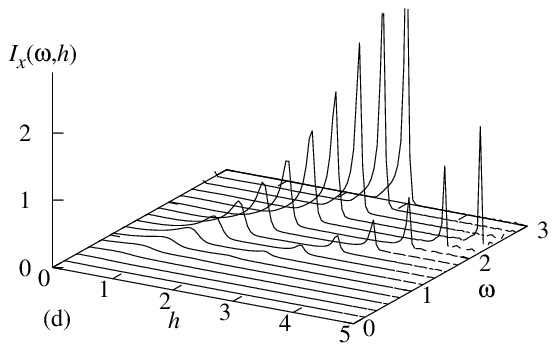}
\caption
{Field profiles of the absorption intensity $I_{x}(\omega,h)$ at different frequencies $\omega$ for $J=-1$, $g_1=1$,
$g_2=1$ (a),
$g_2=0.5$ (b),
$g_2=0$ (c),
and
$g_2=-0.5$ (d)
at $T=1$.}
\label{x16}
\end{center}
\end{figure}

In Fig.~\ref{x16} we present the absorption intensity $I_x(\omega,h)$ as a function of the magnetic field.
In contrast to the $I_{z}(\omega, h)$ case, here the field profiles do not exhibit any singularities.
A prominent feature of the absorption profiles $I_x(\omega,h)$ is a two-peak structure for the case of different nonzero $g$-factors.
The cases $g_2=0.5$ and $g_2=-0.5$ demonstrate additional satellite peak [Figs.~\ref{x16}(b,d)].
For the uniform chain ($g_1=g_2$) we can see one peak which moves with increasing of frequency to a higher value of magnetic field [Fig.~\ref{x16}(a)].
Qualitatively the same picture is seen for $g_2=0$ in Fig.~\ref{x16}(c), where the peak is less steeper in comparison to the case in Fig.~\ref{x16}(a).

\section{Summary}
\label{sec06}
\setcounter{equation}{0}

To summarize,
we have studied the effect of the alternation of $g$-factors
on the static and dynamic properties of the spin-1/2 $XX$ chain in a transverse field.
The crucial point is that the conservation of the total magnetization is lost in this case.
This evokes non-trivial changes in the thermodynamic and dynamic behavior of the model.

While the logarithmic peculiarities of the magnetization and the susceptibility at $T=0$ were obtained earlier \cite{Kontorovich1968},
we found peculiarities in the low-temperature thermodynamics.
In particular,
we have shown that the specific heat can change its behavior
from the linear dependence in the spin-liquid phase to the $\sqrt{T}$ dependence at the saturation field,
and finally transformed to the exponential law (\ref{408}).
The susceptibility at zero magnetic field displays the logarithmic divergence with temperature as it follows in Eq.~(\ref{chi_h=0}).

We have performed the detailed study of the dynamic properties.
We calculated the dynamic structure factors $S_{zz}(\kappa,\omega)$ and $S_{xx}(\kappa,\omega)$
and inspected how they change in the external magnetic field for different period-2 alternations of $g$-factors.
In the case when both $g$-factors are of the same sign,
the correspondence between the boundaries of the $zz$ and $xx$ structure factors is still present
like it was observed previously \cite{Derzhko2000, Derzhko2002}.
On the contrary, if $g_1 g_2 \leq 0$,
a large enough magnetic field leads to the highly intense modes in the $xx$ structure factor.
In addition,
we calculated the absorption intensity $I_{\alpha}(\omega,h)$ for the different configuration of ESR experiments.
In the Voigt configuration ($\alpha=z$),
the model with uniform $g$-factors does not have any response.
In the case when $g_2$ differs from $g_1$, we obtain the nonzero contribution to the absorption intensity.
For sufficiently large frequencies $\omega>2|J|$ the van Hove singularity arises at $h=\sqrt{\omega^2-4J^2}/(2|g_{-}|)$.
In the Faraday configuration ($\alpha=x$),
the situation is a bit different.
The absorption spectra can be observed in the uniform case.
It shows a broad maximum at some resonance field.
The alternation of $g$-factor leads to the doubling of this resonance line.
Although in our study we focus on the exactly solvable $XX$ chain,
from Ref.~\cite{Mueller1981} we know that such analysis of dynamics
is useful for understanding a more realistic case of the Heisenberg chains.

\section*{Acknowledgments}

The present study was supported by the ICTP (OEA, network-68 and NT-04):
V.~O. acknowledges the kind hospitality of the ICMP during his visits in 2015--2019;
T.~V. and O.~B. acknowledge the kind hospitality of the Yerevan University in 2016, 2017, and 2018.
The work of T.~K. and O.~D. was partially supported
by Project FF-30F (No.~0116U001539) from the Ministry of Education and Science of Ukraine.
V.~O. acknowledges the partial support from the ANSEF project condmatth-5212,
as well as the support from the HORIZON 2020 RISE "CoExAN" project (GA644076).

\section*{Appendix: Boundaries of the two-fermion excitation continua}
\renewcommand{\theequation}{A\arabic{equation}}
\setcounter{equation}{0}

Let us present the expressions for the lines in the $(\kappa,\omega)$ plane,
which restrict the regions for different number of solutions of Eqs.~(\ref{506}) as it is shown in Fig.~\ref{x05}(a); green lines.
We have
\begin{eqnarray}
\label{a01}
&&\omega_{1,2}(\kappa)
=
\sqrt{2\left(J^2+2g_-^2h^2\pm J^2\cos\kappa\right)},
\\
&&\omega_{3,4}(\kappa)
=
\left\vert\sin\kappa\right\vert
\left(\sqrt{J^2+g_-^2h^2} \pm \left\vert{g_-h}\right\vert\right),
\nonumber\\
&&\omega_{5,6}(\kappa)
=
\sqrt{J^2\sin^2\kappa+{g_-^2h^2}}\pm\left\vert{g_-h}\right\vert.
\nonumber
\end{eqnarray}

Let us also present the expressions in the case $|h|<h_s$, $g_1 g_2>0$ for the characteristic lines,
which bounded nonzero values of the Fermi-Dirac functions at $T=0$ [see also Fig.~\ref{x05}(c); red lines].
We have
\begin{eqnarray}
\label{a02}
\omega_{7,8}(\kappa)
&=&
\left\vert
{g_+ h} {+} \sqrt{J^2\cos^2\left(\kappa_0\pm\kappa\right){+}{g_-^2h^2}}
\right\vert,
\\
\omega_{9,10}(\kappa)
&=&
\left\vert
{g_+ h} {-} \sqrt{J^2\cos^2\left(\kappa_0\pm\kappa\right){+}{g_-^2h^2}}
\right\vert.
\nonumber
\end{eqnarray}
Here $\kappa_0$ is defined in Eq.~\eqref{302}.

\end{document}